%% file: main.tex
\documentclass[conference,table]{IEEEtran}

\IEEEoverridecommandlockouts

\usepackage[english]{babel}
\usepackage[letterpaper,top=2cm,bottom=2cm,left=3cm,right=3cm,marginparwidth=2cm]{geometry}

\usepackage{amsmath,amssymb,amsfonts}
\usepackage{graphicx}
\usepackage[colorlinks=true, allcolors=blue]{hyperref}
\usepackage{booktabs}
\usepackage{adjustbox}
\usepackage{cleveref}
\usepackage{todonotes}
\usepackage{array} 

\usepackage[utf8]{inputenc}
\usepackage[T1]{fontenc}

\usepackage{subfig}

\usepackage{balance}

\title{U Can't Gen This?\\
A Survey of Intellectual Property Protection Methods for Data in Generative AI}

\author{
    \IEEEauthorblockN{Tanja Šarčević}
    \IEEEauthorblockA{\textit{SBA Research},
        Vienna, Austria\\
        tsarcevic@sba-research.org
    }
\and
    \IEEEauthorblockN{Alicja Karlowicz}
    \IEEEauthorblockA{\textit{SBA Research},
        Vienna, Austria\\
        akarlowicz@sba-research.org
    }
\and
    \IEEEauthorblockN{Rudolf Mayer}
    \IEEEauthorblockA{\textit{SBA Research},
        Vienna, Austria\\
        rmayer@sba-research.org
    }
\and
    \IEEEauthorblockN{Ricardo Baeza-Yates}
    \IEEEauthorblockA{\textit{EAI, Northeastern University},
        Silicon Valley, CA, USA\\
        rbaeza@acm.org
    }
\and
    \IEEEauthorblockN{Andreas Rauber}
    \IEEEauthorblockA{\textit{TU Wien},
        Vienna, Austria\\
        andreas.rauber@tuwien.ac.at
    }
}

\begin{document}
\maketitle
\thispagestyle{plain}
\pagestyle{plain}

\begin{abstract}
Large Generative AI (GAI) models have the unparalleled ability to generate text, images, audio, and other forms of media that are increasingly indistinguishable from human-generated content.
As these models often train on publicly available data, including copyrighted materials, art and other creative works, they inadvertently risk violating copyright and misappropriation of intellectual property (IP). Due to the rapid development of generative AI technology and pressing ethical considerations from stakeholders, protective mechanisms and techniques are emerging at a high pace but lack systematisation.

In this paper, we study the concerns regarding the intellectual property rights of training data and specifically focus on the properties of generative models that enable misuse leading to potential IP violations. Then we propose a taxonomy that leads to a systematic review of technical solutions for safeguarding the data from intellectual property violations in GAI.

\end{abstract}

\section{Introduction}

\begin{figure}[ht]
    \centering
    \includegraphics[width=\linewidth]{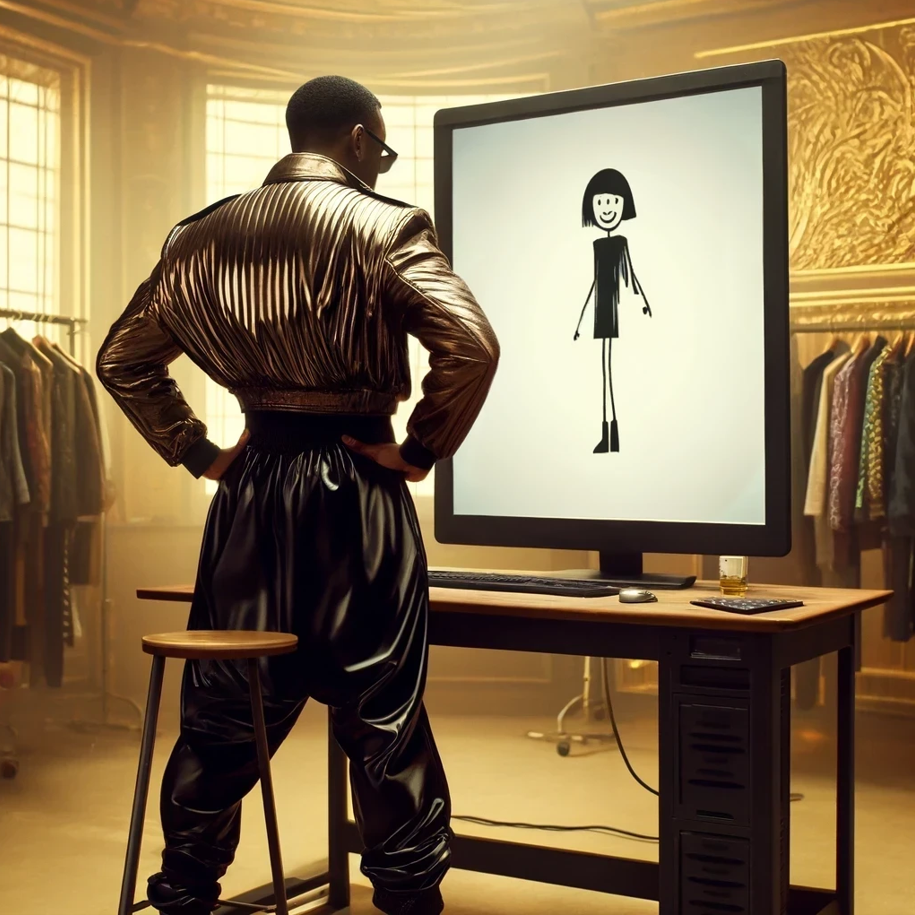}
    \caption[]{\textit{U can't gen this? MC Hammer failing to generate a photo of himself}" 
    \footnotemark.
    } 
    \label{fig:introduction}
\end{figure}

Generative Artificial Intelligence (GAI) has revolutionised various domains, demonstrating remarkable capabilities in generating realistic text~(ChatGPT~\cite{openai_gpt-4_2023}), music~(MuseNet\footnote{\url{https://openai.com/research/musenet}}) and images~\cite{rombach_high-resolution_2022,ramesh_hierarchical_2022} (see \Cref{fig:introduction}), among other outputs.
This cutting-edge technology has the potential to disrupt traditional creative processes, enabling AI systems to autonomously produce content that mirrors human creativity. 
While GAI opens up exciting opportunities for innovation and artistic expression, it also presents significant challenges concerning intellectual property rights and ethical considerations.
\footnotetext{Generated using DALL-E 3, \url{https://openai.com/dall-e-3}}
Intellectual property (IP) rights encompass a set of legal protections designed to safeguard the creations of human intellect. These rights, including copyright, patents and trademarks have played a vital role in promoting creativity, innovation, and fair compensation for creators. However, the emergence of GAI has sparked debates surrounding the nature of these rights when AI systems generate original content without direct human intervention but rather by generalising from a large corpus of human creation.

Recently, artists have been noticing the evidence of the non-originality of generative models through some controversial examples. 
The name of a Polish artist, Greg Rutkowski, known for his fantasy illustrations, appeared in more than 93,000 prompts of Stable Diffusion - the prompts have been used to create art in his distinct style, very successfully so. 
As a result, many AI-created images inspired by Rutkowski's style have been published and appear in online searches~\cite{heikkila_this_2022}. The case was followed by more artists reporting the same behaviour of generative models. 
Three artists, Sarah Anderson, Kelly McKernan, and Karla Ortiz filed a class action lawsuit against Stability AI\footnote{\url{https://stability.ai/}}, Midjourney\footnote{\url{https://www.midjourney.com/}}, and Deviant Art\footnote{\url{https://www.deviantart.com/}}, over their usage of Stable Diffusion~\cite{rombach_high-resolution_2022}, alleging copyright infringement. They assert that these models used their works and the works of thousands of other artists as training data without authorisation.
Another lawsuit was filed by Getty Images against Stability AI for using more than 12 million images and associated metadata without authorisation or compensation to train their model~\cite{nolan_ai_2023}.

One major aspect of the underlying problem and an aspect that distinguishes these current cases from historic cases of plagiarism and other forms of copyright infringement is the sheer scale.
While the artists take years of education and practice to develop their distinct artistic style and ultimately profit from commissioning their art, tools that use their art without consent and create a competitive alternative in an instant, seemingly without a quantitative limit, present a viable threat to their work and careers.

This situation sparked research on mitigation methods for potential intellectual property rights violations in GAI. 
The protection encompasses a diverse array of technical strategies, each tailored to address specific vulnerabilities.
For instance, some techniques aim to support the artist in opting out of large scraped datasets, while other works focus on disabling style mimicry behaviours. 
Many of these techniques build upon prior research on other ethical issues in GAI, such as privacy~\cite{matsumoto_membership_2023}, or inappropriate content generation~\cite{gandikota_erasing_2023}.

We have now arrived at a point where a multitude of technical solutions are available, but a systematisation of the threats and vulnerabilities they address, their costs (e.g., in terms of distorting original artwork), and their shortcomings for protection is lacking.
This paper addresses this gap. It systematically reviews the current state-of-the-art technical approaches for safeguarding training data against IP violations. 
We outline the potential violations of intellectual property considered in the literature and provide a discussion about the underlying properties of generative models that enable such violations. 
Following this, we present a variety of technical strategies, ranging from cleaning training data, model modifications, adversarial injections and policy-driven frameworks, that have been proposed to address these concerns. 

Our contributions include:
\begin{itemize}
    \item a review of potential IP violations of training data for GAI models;
    \item a systematic review of technical solutions for protecting the IP of content available for training the large GAI models; 
    \item a taxonomy and classification of IP protection methods; and
    \item discussion on policy and practice revolving IP issues in the context of GAI.
\end{itemize}

The remainder of this paper is organised as follows: \Cref{sec:related-work} discusses other survey papers on the intersection of intellectual property rights and GAI. In \Cref{sec:methodology}, we describe our research methodology. In \Cref{sec:background}, we provide a detailed background on the technology underlying large generative models, while in \Cref{sec:IPR-threats}, we discuss the threats to IP rights in GAI applications. In \Cref{sec:mitigation} we provide a taxonomy and a systematisation of the state-of-the-art on protecting the IP of training data. 
In \Cref{sec:discussion} we discuss the surrounding landscape of issues in GAI and IP protection. Finally, we provide conclusions in \Cref{sec:conclusion}.

\section{Related work}\label{sec:related-work}
Several works have investigated various issues surrounding IP rights in the age of generative AI, from different perspectives.
Works considering more general issues of GAI, such as privacy and security of AI-generated content (AIGC), mention and describe methods for IP protection from the technical perspective -- however, do not provide sufficient detail. For example, Wang et al.~\cite{wang_survey_2023} cover only watermarking methods when it comes to IP protection in GAI. Chen et al., in two consecutive surveys~\cite{chen_challenges_2023,chen_pathway_2023}, outline issues such as misinformation, toxicity, the vulnerability of models due to malicious usage, and copyrights of AIGC, but do not discuss and classify technical protection approaches in-depth.
Wang et al.~\cite{wang_security_2023} address several relevant methods. For example, they consider \textit{concept removal} to remove toxic content, while we, however, consider the angle of protecting the IP of training data.
Zhong et al.~\cite{zhong_copyright_2023} provide an empirical review that evaluates three methods for IP protection of GANs: adversarial-based, watermarking, and attribution. While they demonstrate their effectiveness for copyright protection, they do not include a comprehensive review and classification of existing IP protection approaches. 

Several papers provide a survey on the legal~\cite{smits_generative_2022,chesterman_good_2023} and societal~\cite{hristov_artificial_2019} perspective of IP rights in the context of generative AI, but do not discuss technical protection methods. 

There are several surveys on underlying concepts of IP protection in generative AI. We discuss in \Cref{sec:mitigation-adversarial} that many methods rely on the creation of adversarial examples and poisoning attacks. 
For discriminative deep neural networks, Zhang et al.~\cite{zhang_adversarial_2019} systematise work on adversarial examples while Cina et al.~\cite{cina_wild_2023} summarise the poisoning attacks and defences.
Sun et al.~\cite{sun_adversarial_2023} focus on adversarial attacks against generative models (GANs and VAEs).
Although those methods are considered attacks on neural networks, they are used in IP protection as they show an effective way of manipulating the model towards a certain behaviour. 
This inverted scenario has already been investigated for IP protection of ML models, where model watermarking is achieved via adversarial examples, discussed in surveys on IP protection of ML models~\cite{lederer_identifying_2023,regazzoni_protecting_2021}.

\section{Methodology}\label{sec:methodology}
\begin{figure}
    \centering
    \includegraphics[width=\linewidth]{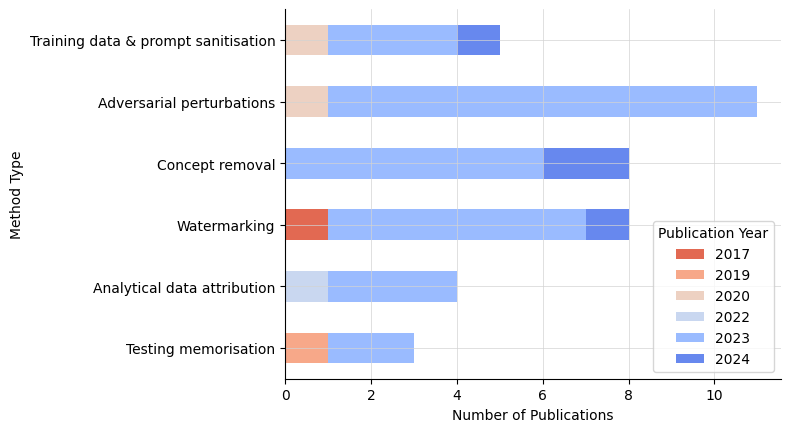}
    \caption{The literature distribution over types of protection methods and year of publication.}
    \label{fig:stats}
\end{figure}
This paper focuses on technical approaches for safeguarding the intellectual property of data used to train the generative models, such as LLMs, GANs and diffusion models. 
Issues relating to the intellectual property and copyrights of training data, in large part, intersect with related ethical issues such as privacy and the generation of harmful content. Therefore, many solutions use similar ideas to achieve diverse goals, often with subtle differences in how the techniques need to be used. 
We hence focus our scope on the methods that directly evaluate at least one IP violation scenario, and exclude other works; however, in \Cref{sec:discussion}, we compare the studied methods to, e.g., methods for removing harmful content.

Our literature research is based on systematic literature studies principles, following the guidelines by Kitchenham et al. \cite{kitchenham_guidelines_2007} and Wohlin \cite{wohlin_guidelines_2014} to generate the collection of peer-reviewed conference papers and journal articles (\textit{formal literature}), as well as pre-prints of methods that in quality are close to peer-reviewed papers, and also provide open-source code of their solutions (\textit{grey literature}). 
We also consider other forms of grey literature, such as blogs and articles, if they serve as technical reports of highly relevant techniques.  

The main reason for the inclusion of grey literature is the high pace of the field and the fact that many influential works might be currently under review at formal publishing venues.
\Cref{fig:stats} shows the distribution of the publications of protection methods and their publication year. 
Overall, we can see that the vast majority of relevant literature on technical solutions to protect IP rights appeared just within the last year -- there are only five papers published before 2022, but 35 papers since.

When analysing the different focal points of the technical approaches, shown in \Cref{fig:stats}, we can see that the most studied approach is \textit{adversarial perturbations}, with eleven papers studying them in the context of IP protection in generative AI.
A further eight papers each consider \textit{watermarking} and \textit{concept removal}, while \textit{training data and prompt sanitisation} is investigated by five works.
The least researched techniques are \textit{analytical data attribution} with four papers and \textit{testing memorisation} with 3 papers.

\section{Background}\label{sec:background}
In this section, we describe the required background and state-of-the-art of large GAI models.

\subsection{Variational Autoencoders}
Variational Autoencoders (VAEs)~\cite{kingma_auto-encoding_2022} are a type of probabilistic generative model composed of an encoder and decoder trained together. The encoder transforms the input to a lower-dimensional latent space, while the decoder reconstructs the input from this compressed representation. Unlike conventional Autoencoders, which learn a deterministic mapping from the input to the latent space, VAEs introduce probabilistic modelling, as the input space is mapped to a probability distribution. This way, VAEs express not just single points in the latent space, but distributions of potential representations. Because the latent space is continuous, the decoder can generate novel data points that smoothly interpolate among training data points. 
VAEs are learned using both a reconstruction loss, which minimises the dissimilarity between original and reconstructed data, and a regularisation loss. The regularisation loss, in the form of the Kullback-Leibler divergence, ensures a balanced distribution of latent representations, thus preventing overfitting and encouraging the encoder to map the latent space close to a Gaussian distribution. 
VAEs are flexible in the type of data they process and can be combined with other generative models~\cite{bao_cvae-gan_2017, pandey_diffusevae_2022}.

\subsection{Generative Adversarial Networks}
Generative Adversarial Networks (GANs), introduced by Goodfellow et al.~\cite{goodfellow_generative_2014} in 2014, consist of two competing networks: the generator and the discriminator. 
The generator generates novel output from the data distribution, by transforming sampled noise vectors. During the training process, the generator learns the mapping of latent space (sampled vectors) to the target data distribution.
Simultaneously, a discriminator network is trained to differentiate between samples generated by the generator (fake samples) and original inputs (real samples). 
The generator, in response, adjusts its parameters to fool the discriminator into accepting its outputs as authentic representations, improving its ability to produce realistic samples. 
Through this adversarial training, both networks compete with each other to strike a balance between their performance and produce high-quality samples. 

The major body of research has focused on GANs for the image domain, and since the introduction of the first works, many variations and implementation improvements have emerged resulting in state-of-the-art architectures such as ProGAN~\cite{karras_progressive_2018}, BigGAN~\cite{brock_large_2019} and StyleGAN~\cite{karras_style-based_2019} for generation task, or CycleGAN~\cite{zhu_unpaired_2020} and StarGAN~\cite{choi_stargan_2018} for image-to-image translation (translating an image from source domain to target domain; I2I).

\subsection{Diffusion Models}
Diffusion models, inspired by non-equilibrium thermodynamics, are based on iteratively destroying an input image and training a network to reverse this process. The concept dates back to 2015 when Sohl-Dickstein introduced the baseline idea~\cite{sohl-dickstein_deep_2015}. However, it was not until Ho et al.~\cite{ho_denoising_2020} and Dhariwal and Nichol~\cite{nichol_improved_2021} improved upon the architecture and implementation that diffusion models gained widespread recognition, later proving their utility and advantages over GANs~\cite{dhariwal_diffusion_2021}.
In the forward diffusion process, the model gradually adds Gaussian noise to the input image until it becomes pure noise. Conversely, in the reverse process, the model predicts and removes the added noise in a similar, gradual way. At inference (generation) time, the model samples from a Gaussian distribution and performs denoising to generate a novel image.

Diffusion models suffer from high computational complexity and long inference time. Latent diffusion models (LDMs) address this issue by operating in the latent space rather than directly in the pixel space~\cite{rombach_high-resolution_2022}. Transforming the images into latent representations compresses the data into lower-dimensional forms, which are perceptually equivalent, yet they reduce the complexity of the input and preserve the details in the decoded images.
In LDMs, the encoding and decoding processes are handled separately from the diffusion process. An independent autoencoder is utilised for this purpose -- it is trained to transition between latent representations and images using perceptual loss to compare high-level features and patch-based adversarial objectives to capture finer details and enhance realism. 
Additionally, a regularisation term is added, similar to standard VAEs, in order to prevent high-variance latent representations, by pushing them towards Gaussian distribution. 
Consequently, during the inference phase, LDMs generate a latent representation, which is then translated into a high-quality image by the decoder.

Generation via diffusion models can be conditioned on image labels or complex text prompts semantically describing the desired output image. Text-to-image models utilise text encoders, such as Contrastive Language-Image Pre-Training (CLIP)~\cite{radford_learning_2021}, to capture the natural language of text prompts and map it to embeddings later injected into the model.

\subsection{Large Language Models}
While the models described above are mostly used for image generation, the rise of GAI fuelled by Large Language Models (LLMs) has opened new horizons for text generation. LLMs are general-purpose language models, trained on extensive text datasets crawled from the Internet and other sources. They typically rely on self-attention and transformer-based architectures~\cite{vaswani_attention_2017} to predict the next token in a sequence; they allow capturing long-context dependencies in text. The interaction with LLMs is commonly done through prompt engineering, where users provide specific instructions to the model. LLMs are versatile and can be used for a variety of tasks such as text generation, translation, summarising, and question-answering. Notable models include the family of Generative Pre-trained Transformers (GPTs) from OpenAI~\cite{radford_improving_nodate, openai_gpt-4_2023}, which also features in the popular chatbot ChatGPT. 
Recent advancements in LLMs encompass diverse models such as the Gemini model family~\cite{gemini_team_gemini_2023}, Large Language Model Meta AI (LLaMA)~\cite{touvron_llama_2023}, and StableLM\footnote{\url{https://stability.ai/stable-lm}}, developed by Stability AI.

\section{Threats to the IP of training data in GAI}\label{sec:IPR-threats}
In this section, we first enumerate the potential IP violations of data used to train generative models. 
As this data often includes valuable content such as artworks and journal articles that reflect human creativity and skill, using this content in generative processes sparked a lot of public backlash and lawsuits. In the following, we use these cases to give context and background to each specific threat to IP. 
Secondly, we dedicate a section to the exploitable properties of the generative models that enable IP violations, \textit{memorisation} and \textit{content-style disentanglement}.

\subsection{Potential IP violations}
Intellectual property rights violations in GAI models can occur in various ways. Some common violations include the following.

\subsubsection{Unauthorised data usage}
If a GAI model uses data without proper authorisation, it can lead to violations of data ownership rights. For instance, a group of developers filed a class action lawsuit against Github claiming a violation of the Digital Millennium Copyright Act for unauthorised use of code to develop the Github Copilot\footnote{\url{https://github.com/features/copilot}}. They claim that Github did not comply with open-source licensing terms and appropriate attribution~\cite{losio_first_2022}.

There are some special cases in the scope of a GAI process that describe special types of unauthorised usages:
        \paragraph{Unauthorised training} The AI provider uses the data for training the GAI model without authorisation, or a user uses the data for fine-tuning the models without authorisation.
        \paragraph{Unauthorised editing} Altering the original content without authorisation may be considered copyright infringement. 
        Text-to-image diffusion models and tools such as Instruct-Pix2Pix~\cite{brooks_instructpix2pix_2023}, HIVE~\cite{zhang_hive_2023}, Blended Diffusion~\cite{avrahami_blended_2022} and Imagic~\cite{kawar_imagic_2023} have the capability to edit images based on used instructions, allowing changes to be made to a source image, so-called image-to-image translation (I2I). 
        The convenience of these methods in terms of e.g. skills required over image-editing tools such as Photoshop raises concerns over the potential misuse, including copyright infringement. 
        Artists are exposed to issues from unauthorised edits of their work such as misinterpretation by the audience and financial losses due to alterations that compromise the original ideas and artistic value. 

\subsubsection{Plagiarism and imitation} 
If a GAI model generates content that closely resembles copyrighted works, it may be considered plagiarism and infringe upon the original creator's copyright. 
For example, The New York Times filed a lawsuit against OpenAI, showing many examples in which OpenAI software re-created New York Times stories nearly verbatim~\cite{grynbaum_new_2023}. 
In another example, a few artists have filed lawsuits for generative models copying their styles~\cite{brittain_judge_2023}.
For visual content, an instance of plagiarism or imitation is \textit{style mimicry}~\cite{gal_image_2022}; for text these include \textit{text paraphrasing}, \textit{replicating verbatim} and \textit{idea replication}~\cite{lee_language_2023}, while for audio, the literature mentions \textit{voice cloning}~\cite{qian_autovc_2019}.

\paragraph{Style mimicry in visual art}
\begin{figure}[ht]
    \centering
    \includegraphics[width=\linewidth]{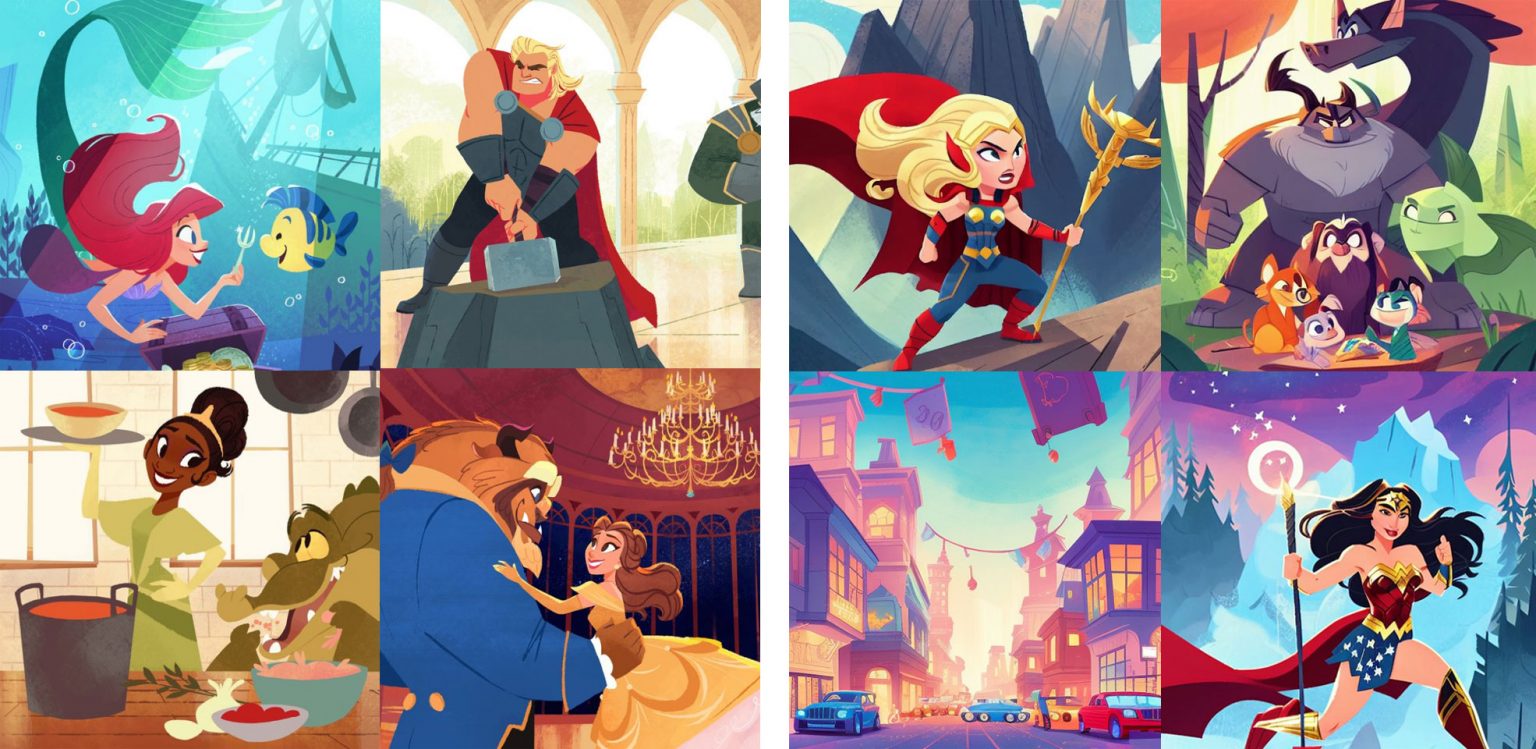}
    \caption{\textbf{Style mimicry} in Stable Diffusion. Left: original artwork by Hollie Mengert vs. Right: images generated in her style~\cite{baio_invasive_2022}.}
    \label{fig:style-mimicry}
\end{figure}
Recently, several artists have found that their works are contained in the training data of large diffusion models and that an end-user is able to generate art in the styles of these artists~\cite{heikkila_this_2022}, initiating numerous copyright lawsuits~\cite{nolan_ai_2023}.
In these cases, the art pieces are not blatantly copied from the training set, however, the art is rather mimicked and appears to be a piece from these artists.
For contemporary artists who are recognised for their distinct style and who are still producing and selling art, such capabilities appear threatening.  

This phenomenon has been called \textit{style mimicry}. The end-user uses an AI model to generate art in a victim artist's style, without the artist's consent. 
This type of attack is mostly prominent in text-to-image (T2I) models where the end-user may prompt text such as "\textit{mountain castle in the style of Monet}" and obtain a high-resolution image of the prompted object (mountain castle) exhibiting distinctive features of Monet's paintings (pastel colours, short brushstrokes and other impressionistic elements). 
These generative models require that the works of an artist are well represented in the training set. This is hence not bound to the great artist from the past, but can also replicate the style of contemporary, working artists, exemplified in \Cref{fig:style-mimicry} for images generated in the style of the artist Hollie Mengert.
 
Mimicry can be achieved for styles that are not necessarily part of the training data for publicly accessible generative models. For this, the attacker needs access to the weights of a well-trained generic text-to-image model and access to artwork samples from the target artist to fine-tune the generic model using the additional samples. 
This is called personalised or customised text-to-image generation (CT2I).
In other words, the end-user can calibrate a generic model to produce images in the style of the target artist even if their art does not initially appear in the training data. 
State-of-the-art CT2I approaches include Textual Inversion~\cite{gal_image_2022} which adds a special text embedding for the new unseen concept, DreamBooth~\cite{ruiz_dreambooth_2023} which fine-tunes a diffusion model on a small number of samples on the same subject and Low-Rank Adaptation (LoRA)~\cite{hu_lora_2021} which modifies a small number of additional weights and adds them to the original parameters of a diffusion model. 

\paragraph{Text plagiarism}
LLMs are trained on vast datasets that include a wide range of sources from the internet, books, articles, etc. 
When generating text, these models may produce content similar or identical to the original training data~\cite{carlini_extracting_2021,mccoy_how_2023}.
Lee et al. identified three types of text plagiarism~\cite{lee_language_2023}:
\begin{itemize}
    \item \textit{replicating verbatim} -- exact copies of words
    \item \textit{paraphrasing} -- synonymous substitution, word reordering, back translation 
    \item \textit{idea replication} -- a representation of the core content in an extended form
\end{itemize}

\paragraph{Voice cloning} 
Voice cloning has been used to create synthetic voices in different industries (voice assistants, video games, speaking aids, etc.).
Voice cloning is achieved either by voice conversion (a.k.a., voice style transfer) or text-to-speech (TTS) generation.
Voice conversion is a technique for modifying the source to a target speaker's speech without changing linguistic information.  
The state-of-the-art voice conversion techniques are typically relying on GANs~\cite{kaneko_cyclegan-vc2_2019} and autoencoders~\cite{qian_autovc_2019}. 
TTS techniques are more complex systems consisting of text analysis, acoustic model and audio synthesis~\cite{wang_tacotron_2017,ren_fastspeech_2022}. 
They exhibit more flexibility in voice generation as they transfer any piece of text to speech without the original speech as a reference.

Voice cloning technology, however, imposes security and privacy threats when used with malicious intentions.
A voice cloning attack is an unauthorised synthesis of a person's timbre (e.g. patterns, intonation and other speech characteristics) with the goal of impersonation or spreading misinformation. 
Thus, voice cloning can cause financial losses, public panic, reputation damage and copyright infringement~\cite{brewster_fraudsters_2021}.

\subsubsection{Reproduction of copyrighted materials} 
    \begin{figure*}[ht]
    \centering
    \includegraphics[width=0.9\textwidth]{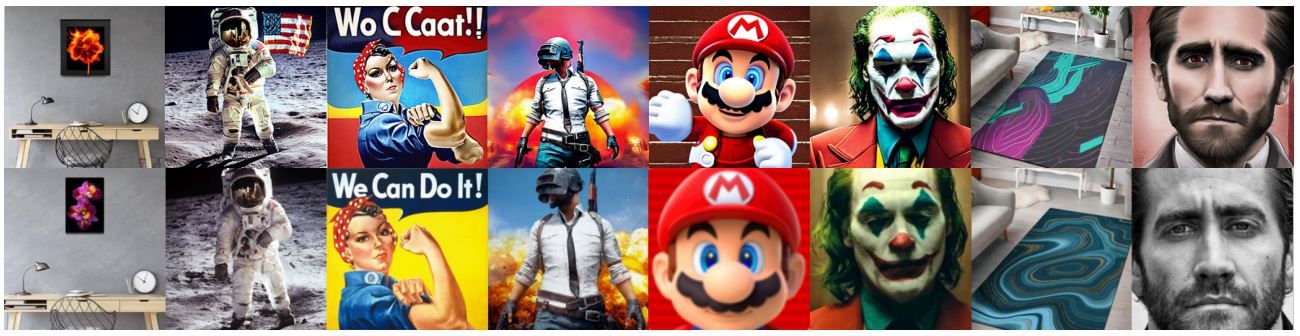}
    \caption{\textbf{Data replication} in Stable Diffusion. Top row: generated images, bottom row: training samples~\cite{somepalli_understanding_2023}.}
    \label{fig:replication-user}
\end{figure*}
If a GAI model reproduces copyrighted images, texts, music, or other creative works without permission, it can violate copyright laws. 
Recent works have shown that state-of-the-art diffusion models can reproduce images from their training data up to near-perfect accuracy when prompting the model with captions directly sampled from the training dataset~\cite{somepalli_diffusion_2022,carlini_extracting_2023}. 
For example, the Stable Diffusion model is capable of reproducing training data as shown in the examples in \Cref{fig:replication-user}.
Furthermore, replication is shown to be possible in some instances even when captions do not mention the content from the generated images. 
This mostly happens via prompts precisely describing the contents of existing works of art~\cite{somepalli_understanding_2023}, hence relating to how an end-user user would create their prompts.

The definition of data replication can be treated differently for different attacker objectives which can span from privacy attacks to IP rights violations. 
For IP rights violations, there needs to be a substantial similarity in the content and its prominent elements between the generated sample and its training sample match.
For the image domain, replicated content is content containing an object, either in the foreground or background, that appears identically in a training image, neglecting minor variations in appearance that could result from data augmentation~\cite{somepalli_diffusion_2022}.
Data replication may occur without the malicious intention of the end-user as well. 
We discuss in \Cref{sec:memorisation} how some generative models are likely to generate copies of the training data samples due to high memorisation capability. 

\subsubsection{Lack of or wrong attribution} 
The problem of attribution in AI-generated relates to determining and acknowledging the original source of content that GAI produces. The process of generating these forms of content complicates the issue of attribution since AI outputs can reflect the influence of numerous original works. 
Real-life scenarios, such as the aforementioned lawsuits against Github and OpenAI, imply and highlight the issues of attribution and compensation to original creators. 
The misattribution issue is particularly problematic because it not only denies creators recognition and compensation but also challenges the very foundations of copyright law, which aims to incentivise creativity and innovation by protecting the rights of the creator~\cite{ulku_kahveci_attribution_2023}.

\subsubsection{Trademark infringement} 
If a GAI model generates content that includes trademarks or logos without proper authorisation, it may infringe upon the rights of the trademark owner. 
Getty Images is alleging that Stable AI, the company behind Stable Diffusion, copied more than 12 million images and associated metadata to build their model without permission from or compensation to Getty Images. 
Moreover, they argue that the generated images have a modified version of the Getty Images watermark attached to odd images, tarnishing the reputation of Getty Images~\cite{vincent_getty_2023}.
    
\subsubsection{Unfair competition} 
If a GAI model is used to create content that imitates a competitor's products or services, leading to confusion or deception among consumers, it can be considered unfair competition. The previously mentioned lawsuit of Getty Images further argues that Stability AI is now a direct competitor to them for creative imagery as it has used a large number of their images. In cases of style mimicry, copyright laws provide limited protection for artistic style, therefore artists stay concerned about the imposed unfair competition.

\subsection{Exploitable properties of generative models}
\subsubsection{Memorisation}\label{sec:memorisation} 
Potential IP violations such as data replication, style mimicry and text plagiarism are possible partially due to the memorisation capabilities of generative models. 
Some works argue that the memorisation in deep learning is an equivalent of overfitting~\cite{feldman_does_2020,feldman_what_2020,arpit_closer_2017}, while others point to local memorisation that occurs simultaneously with generalisation in the same model, making a distinction from global overfitting~\cite{van_den_burg_memorization_2021, tirumala_memorization_2022}. 
In any case, the memorisation capabilities have been shown to exist for a range of generative models for different domains, including specific instances like GANs~\cite{webster_this_2021} and diffusion models~\cite{somepalli_diffusion_2022}, as well as broader category of generative language models~\cite{lee_language_2023,jagielski_measuring_2023,carlini_quantifying_2023}.

Within the domain of image generation, memorisation can be quantified through e.g. image similarity metrics. In the case of near-duplicates, exact memorisation involves detecting matches utilising a straightforward $l_2$ distance computation between the generated image and the training samples in the pixel space~\cite{carlini_extracting_2023}. Alternatively, relative metrics consider the distance of the closest image in relation to other $n$ nearest neighbours in the training set~\cite{carlini_extracting_2023, yoon_diffusion_2023, gu_memorization_2023}.
Although pixel-level memorisation effectively identifies near-copies within the training data, it remains susceptible to deception through minor pixel shifts. Notably, large models such as Stable Diffusion have exhibited the capacity not only to memorise entire images but also to retain partial information, such as individual objects~\cite{somepalli_diffusion_2022}.
For the exploration of object-level memorisation, more complex features are relevant, including embeddings derived from CLIP or features extracted from copy-detection models, like Self-Supervised Descriptor for Image Copy Detection (SSCD)~\cite{radford_learning_2021, pizzi_self-supervised_2022}.

To this date, several, partially contradicting, theories have been formulated on why memorisation occurs in complex, generalised models.
Some works argue that the memorisation capability can be primarily attributed to duplicates in the training sets~\cite{wen_canary_2023}. 
However, it has been shown that de-duplication does not eliminate the memorisation problem ~\cite{somepalli_understanding_2023, carlini_extracting_2023}. 
In the other works, during data extraction, outliers were also shown to be among vulnerable instances~\cite{carlini_extracting_2023, van_den_burg_memorization_2021}.
Other factors that have been found to play an important role in the memorisation behaviour include text conditioning in text-to-image models~\cite{somepalli_understanding_2023}, skewness in the distribution of image repetitions in the training set and the number of gradient updates during training being large enough to enable overfitting on a subset of data~\cite{somepalli_diffusion_2022}. Moreover, conditional diffusion models, where the generation is conditioned on a particular class as opposed to generation from whole training set distribution or guided by text prompt in text-to-image models, demonstrated an increased capacity for memorising images in comparison to their unconditional alternative~\cite{gu_memorization_2023}. 
Furthermore, the dataset size has a crucial impact on the feasibility of replicating images.
This relates to the model's learning capacity, which is influenced by the size of the model and the complexity of the model's architecture. If the dataset size is not big enough for the capacity of the diffusion model, the model will likely fail to generalise. This also applies on a class level, where the model generates novel images from well-represented classes but memorises samples from a minority class~\cite{yoon_diffusion_2023}.

Generally, the works mentioned above~\cite{somepalli_diffusion_2022, somepalli_understanding_2023, carlini_extracting_2023, carlini_extracting_2021} are prone to underestimate the memorisation behaviour of generative models because of computing limitations that would (i) allow comparison with entire training sets of large diffusion models, and (ii) the limitations and inconsistencies of detection methods, i.e. not being able to find some memorised samples due to the missing capability of capturing examples that have semantic relevance, but are not detectable with current simple metrics. 
Large diffusion models are trained on billions of images -- for example, Stable Diffusion uses several versions of the LAION dataset, with the largest of those ranging around 5 billion images\footnote{\url{https://laion.ai/blog/laion-5b/}} -- and hence, finding matches in the entire set is infeasible. To overcome this obstacle, the search is often limited to arbitrary subsets of data, duplicates, outliers, or fine-tuning sets. Further, used detection methods vary between research, depending on the data complexity, sometimes using arbitrary thresholds for match candidates. This poses a challenge to what we count as memorisation and how to quantify and compare it across models and data.
Regardless, trends can be observed between different types of models. Carlini et al.~\cite{carlini_extracting_2023} and Akbar et al.~\cite{akbar_beware_2023} show that GANs are, for instance, substantially less prone to memorisation compared to diffusion models.
This might be because the generators of GANs are only trained using a very high-level abstraction of training data (i.e. using gradients from the discriminator) and, therefore, never receive training data as input, unlike diffusion models, which directly process noised training images many times through training.

An interesting phenomenon is that the set of inputs regenerated by GANs and the set regenerated by diffusion models greatly overlap, i.e. a large portion of the memorised train images are memorised by both types of models. This may 
indicate that the mitigation strategies should be, at least partially, directed towards preprocessing the training set.

Generative language models also exhibit memorisation behaviour. Particularly, the capacity of LLMs to memorise is an active area of research, focused on exploring methods for recovering texts present in the training corpus or techniques to audit the models by extracting artificially injected texts called \textit{canaries}~\cite{carlini_secret_2019}. The replication occurs even in models that are not overfitting to their training data~\cite{tirumala_memorization_2022,jagielski_measuring_2023}. 
The amount of replicated data from language models is generally proportionate to (i) the size of the model, (ii) duplicates in the text data, and (iii) the prompt length~\cite{carlini_quantifying_2023,carlini_extracting_2021}. The information that can be extracted differs from that for images, as from the text data we can obtain personally identifiable information such as names, phone numbers, and email addresses. Additionally, not only single pieces of information can be leaked, but also longer verbatim sequences, like source code and articles~\cite{carlini_extracting_2021}. While extraction commonly focuses on exact text sequences, some studies consider more approximate memorisation aspects, such as algorithms, writing styles~\cite{hartmann_sok_2023} or levels of plagiarism~\cite{lee_language_2023}.

Memorisation is largely studied in the light of \textit{membership inference attacks}, where the adversary's goal is to infer whether a particular data sample was part of the training set for a given model. Membership inference has first been discussed for supervised machine learning models (such as classification or regression models)~\cite{shokri_membership_2017}, but has later been expanded to GAI models as well~\cite{webster_this_2021,hu_membership_2023,matsumoto_membership_2023}.
Another attack exploiting the memorisation capabilities is the \textit{model inversion attack}~\cite{fredrikson_model_2015}, where the objective of the attacker is to recreate one or more (not publicly shared) training samples used to train the target model\cite{yin_dreaming_2020,ghiasi_plug-inversion_2022,carlini_extracting_2023}. 
It is important to note that membership inference can be performed by reconstructing the original training data from the model outputs (i.e. via model inversion), hence both lines of work produce highly relevant insights for intellectual property protection of training data.
In the context of GAI, these attacks build on directed prompt crafting to produce a desired outcome, while violation of IP rights can occur by models spontaneously generating the training set replicas without malicious intentions of the end-user.

\subsubsection{Content-style disentanglement}
A concept in the realm of image generation models closely related to style transfer and mimicry is disentanglement. This property allows for the separation of attributes within the generated image, allowing for targeted modifications of specific aspects like style without changing the image content. Disentangling style from content enables the application of a particular image or artist's style to another image.
Disentanglement is an inherent property of some models such as GANs and diffusion models, which allows for image editing without any additional learning. Disentanglement can be simply done by exploiting the learned latent representation of the generator in a GAN~\cite{collins_editing_2020, harkonen_ganspace_2020} or, for text-to-image diffusion models such as Stable Diffusion, replacing the text embedding with desired modification at later denoising steps during the reverse diffusion process~\cite{wu_uncovering_2023}. This preserves the semantic content of the original image while shifting it toward the target style. The inherent disentanglement capability does not require fine-tuning the model as opposed to the task of disentanglement representation learning~\cite{higgins_towards_2018} for more controlled editing. 
Current works on representation disentanglement learning are mainly focused on unsupervised techniques incorporating encoders and modified loss functions into optimisation such as content-preservation loss and style-disentanglement loss~\cite{yang_disdiff_2023, wang_stylediffusion_2023, kwon_diffusion-based_2023, kazemi_style_2018}.
A widely adopted practice is the incorporation of Adaptive Instance Normalisation (AdaIN)~\cite{huang_arbitrary_2017} layers into the architecture of GANs as it carries the style information of images~\cite{kazemi_style_2018, kwon_diagonal_2021, xu_drb-gan_2021}.
It is worth noting that content-style disentanglement is evaluated for other contexts than artistic styles as well, like attributes of faces or fashion items.

In the context of art style transfer, inherent disentanglement is possible if a representative number of style images is already present in the training data of a model. On the other hand, the specific style of an arbitrary image can be learned, extracted, and applied to desired content through various adaptations of GANs~\cite{kaneko_cyclegan-vc2_2019, liu_artsy-gan_2018, kazemi_style_2018} and diffusion models~\cite{wang_stylediffusion_2023, wu_not_2023}. However, in case of a more complex representation of the artist's style, a collection of images can be used for collection style transfer enabling to capture variations of the artist's work~\cite{kotovenko_content_2019, xu_drb-gan_2021}.

\input{taxonomy}

Recently, Lu et al.~\cite{lu_specialist_2023} demonstrated a method called Specialist Diffusion, a few-shot fine-tuning of Stable Diffusion for content-style disentanglement, specifically designed for synthesising known objects in unknown styles. Here, unknown styles refer to the styles that were not present in the training data. With this method, less than 10 images of any artist suffice to capture their unique aesthetic and create novel images in the same fashion.

\section{IP protection techniques}\label{sec:mitigation}
In this section, we provide an overview and systematisation of state-of-the-art protection techniques for intellectual property rights of training data used for generative models.
\Cref{fig:taxonomy} presents the taxonomy described in \Cref{sec:mitigation-taxonomy} and the categorisation of the protection methods.
These different categories (shown as the leaf nodes of \Cref{fig:taxonomy}) will be further described and discussed in \Cref{sec:mitigation-categories}.
Beyond the technical approaches, protection against IP violation is oftentimes addressed via content policies of e.g. generative AI providers or content publishing sites. 
We discuss these policy-based methods \Cref{sec:mitigation-ethical}.

\subsection{Taxonomy}\label{sec:mitigation-taxonomy}
Our proposed taxonomy is shown \Cref{fig:taxonomy}. We distinguish the methods along their characteristics of the \textit{protection mode}, the \textit{phase of application}, and their \textit{granularity}, as follows:
\subsubsection{Mode of protection}
Mode of protection is the categorisation based on a broader objective towards addressing IP violations. 
It includes:
\begin{itemize}
    \item \textit{Prevention}: methods that include actions taken to mitigate or avoid potential IP violations  
    \item \textit{Detection}: methods that aim to observe and identify potential IP violations after they occur
\end{itemize}
\subsubsection{Phase of application}
Protection methods may be applied at different stages of the GAI lifecycle. 
For \textit{prevention} methods we differentiate:
\begin{itemize}
    \item \textit{Data sharing} phase -- methods are applied to the training data in the sharing phase before they are used for training a GAI model
    \item \textit{Model training/fine-tuning} -- methods are applied during model training or fine-tuning
    \item \textit{Inference} -- methods are applied at inference without interfering with the training process
\end{itemize}
\subsubsection{Granularity}
This categorisation distinguishes between the entities in the GAI process that the method is applied to take effect in protecting the training data. 
This includes:
\begin{itemize}
    \item \textit{Dataset}: these methods focus on modifying the training dataset as a whole to achieve protection
    \item \textit{Sample}: IP protection is achieved by directly applying a method on the sample of interest
    \item \textit{Concept}: IP protection targets certain features representative of a concept, rather than a single input sample. A concept is an abstraction of a set of input samples that belong under the same arbitrary category; an example of a concept in the IP protection context is an artistic style.
    \item \textit{Model}: methods target the model and its behaviour, without interaction with the data, to achieve the IP protection of training data
    \item \textit{Prompt}: methods target the textual prompt, typically modifying it at inference time
\end{itemize}

According to these three levels of categorisation, we identify seven types of protection techniques, as shown in \Cref{fig:taxonomy}. 

\input{terminology}

\begin{itemize}
    \item \textit{Dataset sanitisation} requires direct access to the training set and modifying the underlying data before any training can take place -- with the goal of preventing IP violations further down the AI generation process.
    These techniques are described in \Cref{sec:mitigation-sanitising}, where we also provide a more detailed classification into sub-categories. 
    \item \textit{Adversarial perturbations} as prevention techniques are applied to specific samples that need to be protected, and they cause the targeted behaviour in the downstream generation process. These methods are further described in \Cref{sec:mitigation-adversarial}.
    \item \textit{Unlearning} and \textit{concept removal} methods modify the model parameters and rely on re-training or fine-tuning the model, hence they require access to the training process. These are described in \Cref{sec:mitigation-concept}.
    \item \textit{Watermarking} techniques are proactive detection techniques and are described in \Cref{sec:mitigation-watermarking}.
    \item \textit{Analytical attribution} is one of the reactive detection methods that do not modify the data collection, training or generation process but can rather serve as detection of potential violations, while another reactive detection method, \textit{testing memorisation}, can be used as an indicator that the additional preventive methods should be applied. We provide details on these methods in \Cref{sec:mitigation-detection}.
\end{itemize}

These categories are summarised in \Cref{tab:terminology}. Literature is not coherent in terminology for these protection categories; we, therefore, provide a common nomenclature in this paper.
To this end, we identify terms which are
used synonymously and reference the respective publications.

\subsection{Categorisation of IP protection methods}\label{sec:mitigation-categories}
In the following, we classify the protection mechanisms we identified in our systematic literature search with the help of the taxonomy introduced above (cf. \Cref{fig:taxonomy}).

\subsubsection{Sanitising training dataset and prompt modification}\label{sec:mitigation-sanitising}
These techniques rely on the ability to modify the training data before it is released or used for model training and modifying prompts at inference time.
\paragraph{De-duplicating training data} 
Current models are trained on massive datasets scraped from the internet containing billions of multimedia objects. This naturally includes several duplicates, e.g. if photos are re-used in multiple news articles, and especially also near (approximate) duplicates, which might differ just in metadata (image headers), formats, minute cropping or other non-invasive transformations. 
We discussed in \Cref{sec:memorisation} that duplicated samples (e.g., more than 100 times) are easier to replicate for diffusion models~\cite{carlini_extracting_2023}, GANs and language models~\cite{carlini_extracting_2021,kandpal_large_2023}.
The problem of duplicates is prominent - for the two-billion version of the LAION image dataset ("LAION-2B"), roughly 700 million images are reported to be duplicates~\cite{webster_-duplication_2023}.
De-duplicating is, hence, a mitigation step of removing the exact or approximate duplicates from the training set with the goal of mitigating the memorisation. 

Detecting near-duplicates is challenging, as the search space is big. 
To address this issue, de-duplication can be addressed either by restricting to removing exact duplicates, or by using a proxy measure for near-duplicates, such as matches found with more advanced methods such as CLIP similarity~\cite{radford_learning_2021,webster_-duplication_2023}, i.e. similarity in a latent space. 
Once duplicates are identified, this method is straightforward and can significantly reduce the number of replicas generated at inference time e.g., \cite{carlini_extracting_2023} has shown up to a 23\% decrease in replicas for diffusion models). 
Memorisation, however, cannot be fully alleviated by de-duplication alone as other factors, such as image caption specificity, the size of a dataset and length of training, contribute to the memorisation behaviour~\cite{somepalli_understanding_2023}.

\paragraph{Image caption modification}\label{sec:mitigation-sanitising-caption}

\begin{figure*}
    \centering
  \subfloat[\label{fig:whitebox-workflow}Train time: multiple captions]{%
       \includegraphics[height = 3.8cm]{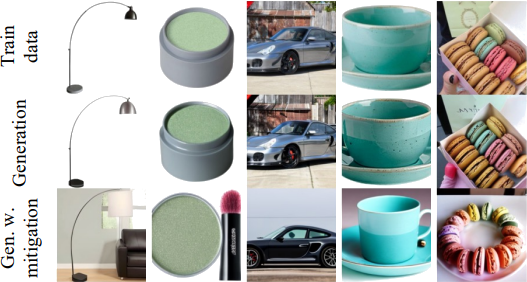}}
    \hfill
  \subfloat[\label{fig:blackbox-workflow}Inference time: random token replacement / addition]{%
        \includegraphics[height = 3.8cm]{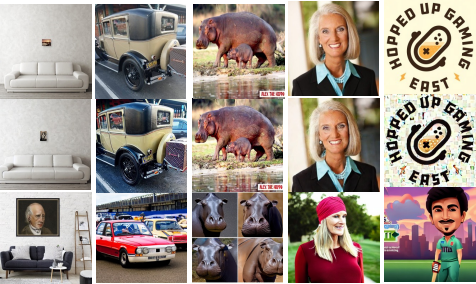}}
    \caption{\textbf{Caption modification}: The most successful mitigation strategies in~\cite{somepalli_understanding_2023} against replication in Stable Diffusion: (a) \textit{multiple captions} for train time and (b) \textit{random token / number addition} for inference time. The top row represents train samples, the middle row is the memorised image, and the bottom row is the generation with a mitigation strategy applied.}
    \label{fig:caption-modification}
\end{figure*}

The memorisation capabilities of text-to-image models are amplified when duplicated training images also come with duplicated image captions. 
By de-duplicating the image captions, the replication of training samples in generative models can be significantly reduced (even if no training image de-duplication has been performed). 

In \cite{somepalli_understanding_2023}, the authors propose several techniques for reducing data replication by randomising and augmenting image captions in the training set. 
These strategies can be applied at train time to increase caption diversity among duplicates, but also at inference time which we discuss in the next paragraph.
These strategies include: 
\begin{itemize}
    \item Multiple captions: for instance, using some of the pre-trained vision-language frameworks for generating synthetic captions (e.g., BLIP~\cite{li_blip_2022})
    \item Adding a small amount of Gaussian noise to text embeddings
    \item Random caption replacement: replacing the existing caption of an image with a random sequence of words
    \item Random token replacement and addition: randomly replace words in the caption with a random word, or add a random word to the caption at a random location
    \item Random word repetition: repeat a random word from a caption at a random location of the same caption
    \item Random numbers addition: add a random number at a random location in the caption
\end{itemize}
While all proposed strategies reduce the similarity of the generated images to the corresponding training sample to some extent, \cite{somepalli_understanding_2023} showed that for Stable Diffusion, the most effective strategy is introducing \textit{multiple captions} at train time, and \textit{random token replacement/addition} at inference time; both strategies are shown in \Cref{fig:caption-modification}.

Another caption modification method was proposed in ~\cite{li_mitigate_2024}. 
In order to mitigate replication, they generalise image captions to reduce specificity, as highly specific captions can act as keys to the memory of the model. 
The generalisation is done by an LLM, specifically, GPT-3.5, guided by the prompt containing instructions for generalisation. 
Additionally, they utilise an additional fusion dataset to sanitise the original data. 
This is done by combining the text embedding of the generalised caption with text embedding from a randomly selected image from the fusion dataset, which is also done with the corresponding image encodings. This augmentation strategy increases the diversity of the data and reportedly reduces the replication rate.

\paragraph{Prompt modification}\label{sec:mitigation-sanitising-prompt}
Post-hoc methods that do not modify the models or data, but rather change the textual prompts in T2I generation fall under \textit{prompt modification} methods.
These generally include rephrasing the textual prompt to increase the diversity of generated content and prompt moderation to prevent specific (e.g. harmful or copyright-violating) content from being generated. 

The strategies for modifying image captions by Somepalli et al.~\cite{somepalli_understanding_2023} mentioned in the previous paragraph are also applied at inference time to the prompt, to disrupt the memorised connection between a caption and an image and hence reduce the replication of training samples. 
For prompt modification at inference time, the most effective strategy is \textit{random token replacement or addition}, shown in \Cref{fig:caption-modification}.
This way, data replication is significantly reduced. 
Another goal is to prevent the generation of certain concepts.
Prompt-checkers at inference time achieve this by identifying and rejecting malicious prompts~\cite{hanu_detoxify_2020,noauthor_openai_2023}.
These methods, hence, require lists of malicious concepts and labelled datasets (often relying on user feedback) to train classification models that effectively classify malicious content. 
OpenAI Moderation~\cite{noauthor_openai_2023}
, for instance, includes categories of malicious content such as hate, harassment and violence and employs a language model to e.g. remove or generalise the names of public figures. 
A challenge of prompt modifying techniques is dealing with unseen malicious content, although extending it is rather simple and does not require retraining the model.
 
\subsubsection{Adversarial perturbations}\label{sec:mitigation-adversarial}
\begin{figure*}[ht]
    \centering
    \includegraphics[width=0.9\textwidth]{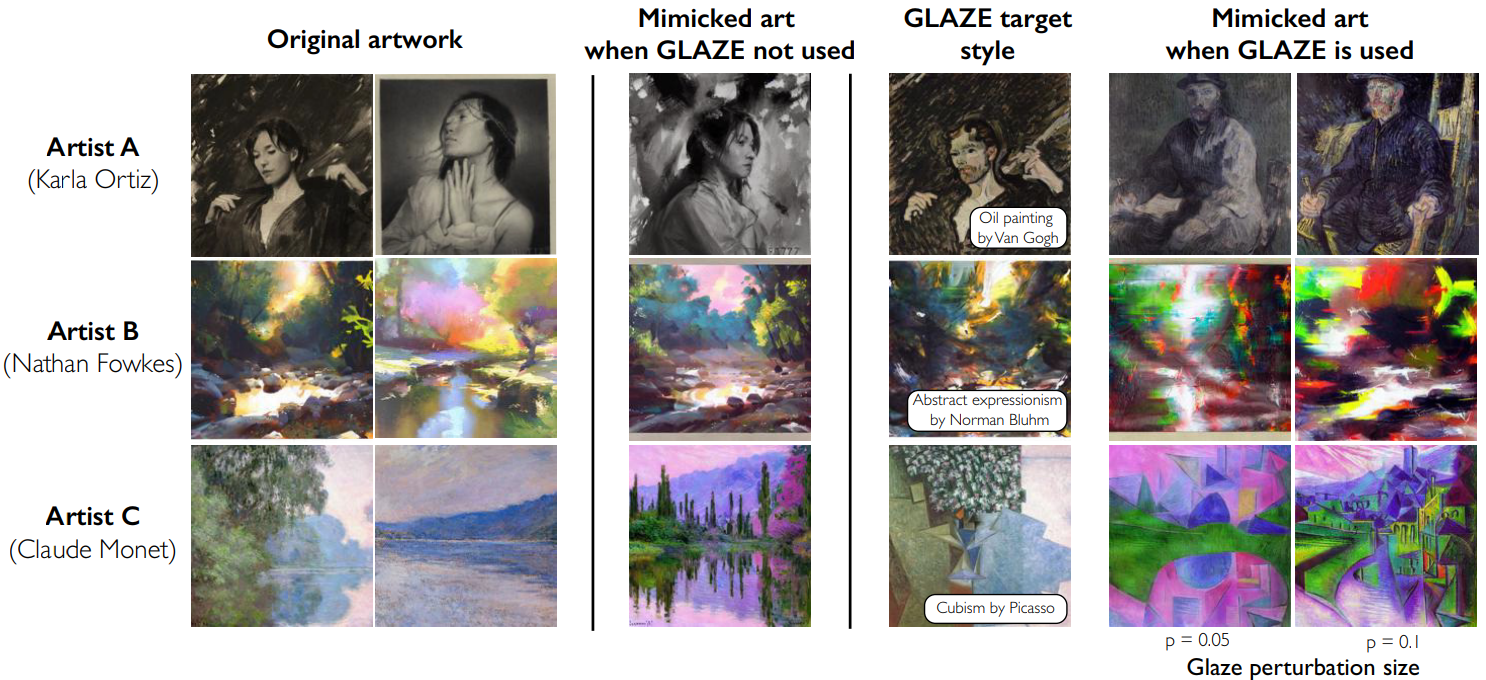}
    \caption{\textbf{Cloaking} via Glaze~\cite{shan_glaze_2023}: columns 1-2 show original artwork from three artists; column 3 depicts results from style mimicry using Stable Diffusion; column 4 is the representation of the target, "cloak" style; columns 5-6 are (failed) attempts of style mimicry based on the cloaked artwork.}
    \label{fig:cloaking}
\end{figure*}
Content owners can actively protect their intellectual property from GAI through a strategic application of adversarial examples. 
Adversarial examples~\cite{goodfellow_explaining_2015} are inputs to machine learning models in the inference stage that are deliberately designed to cause the model to make a mistake. 
They employ changes to input data that are generally imperceptible to humans and target the learned features and decision boundaries, ultimately leading to incorrect outputs at the inference time. 
Drawing the motivation from the well-researched realm of adversarial examples for discriminative models, Liang et al. introduce adversarial examples for generative models~\cite{liang_adversarial_2023} with their method called AdvDM.
Unlike when applied against classification models, where adversarial examples aim to cause misclassification, against diffusion models, adversarial examples disrupt the generation process, by preventing features from being extracted. 
This property of adversarial examples enables a direct application in safeguarding intellectual property in the domain of AI-generated content: employing adversarial perturbations can cause the models to fail to reproduce or learn from the specific style of the protected artworks and/or cause low-quality generated content. 
Across the literature, these modifications have been named differently, due to the rapid concurrent development and the ensuing lack of a standard terminology. For instance, the modifications may be referred to as \textit{adversarial perturbations}~\cite{liang_adversarial_2023,liang_mist_2023}, \textit{adversarial noise}~\cite{zhao_unlearnable_2023}, \textit{watermark}~\cite{ye_duaw_2023}, \textit{style cloak}~\cite{shan_glaze_2023}; the method of applying these modifications has been dubbed \textit{immunisation}~\cite{salman_raising_2023}, \textit{cloaking}~\cite{shan_glaze_2023} or \textit{poisoning}~\cite{liu_metacloak_2024}. 
 
AdvDM~\cite{liang_adversarial_2023} is a method to create these adversarial examples designed against diffusion models and utilises the training loss of these models, referred to as \textit{semantic loss}. 
This semantic loss aims to shift the image representation away from the semantic space of the model, leading to the production of images with chaotic content when generated by the model.
In their subsequent work, Liang and Wu introduce Mist~\cite{liang_mist_2023}, a method that significantly improves transferability and robustness of adversarial examples across different diffusion model applications, by optimising a novel adversarial loss term, \textit{textural loss}\footnote{The authors interchangeably use textural and textual loss for describing the same loss; we believe that the correct term is textural loss as this loss exhibits modifications in \textit{textural} characteristic on the image background}. 
Textural loss represents the distance between the encoded representation of the original image and the representation of the perturbed image in the latent space of latent diffusion models. 
The adversarial example is then generated by adding subtle perturbation to the image to maximise the textural loss. As this approach targets the encoder, it can thus only protect against Latent diffusion models. 
The method of creating Unlearnable Diffusion Perturbation (UDP)~\cite{zhao_unlearnable_2023} further increases the distortions in the generated content in order to deter from usage of these images for training or fine-tuning the diffusion models. 
To achieve the unlearnable effect, the design objective is to increase the distance between the distribution of generated images and the distribution of clean training data as much as possible by adjusting the protective noise.
Based on the observation that the existing adversarial attacks introduce an additional error to the score function predicted by LDMs which in turn leads to biased predictions on these adversarial examples, Zheng et al.~\cite{zheng_understanding_2023} propose Attacking with Consistent score-function Error (ACE).
ACE standardises the pattern of the extra error added to the predicted score function. 
This forces the fine-tuned LDM to learn a consistent pattern as a bias when predicting the score function, contributing to the effectiveness of the method.

Glaze~\cite{shan_glaze_2023} also aims at misleading generative models from accurately mimicking the specific artist's style, by perturbations that are optimised to shift the image representation in latent feature space, similar to the idea of textural loss in Mist.
Glaze, however, targets specifically the style features instead of compromising the quality of the generated output. 
To this end, they first apply style transfer to the original art. 
Style transfer modifies the original image to adopt the appearance of a different style - the target style -- while retaining the content from the original art. 
The perturbation, named style cloak, is then obtained by minimising the distance between the feature representation of the original and the style-transferred image.
As a result, the applied perturbations are not recognisable by humans, but the model will generate a recognisably different style when prompted to generate art in the victim's style. This is illustrated in \Cref{fig:cloaking}. 

Ye et al. propose their Data-free Universal Adversarial Watermark (DUAW)~\cite{ye_duaw_2023}. 
DUAW disrupts the variational autoencoder in Stable Diffusion models, ensuring that images altered by the adversarial perturbation (the authors call this "watermark") produce significantly distorted outputs when used for model customisation (customised text-to-image generation (CT2I)), e.g. with approaches like DreamBooth~\cite{ruiz_dreambooth_2023} or LoRA~\cite{hu_lora_2021}. 
Hence, this method, besides rendering an artistic style unrecognisable, also disrupts the model as a result of the unauthorised usage of protected input. 
Unlike the previous methods, DUAW operates without direct access to the images to be protected, using synthetic data for optimising the adversarial perturbation instead. 
Nightshade~\cite{shan_prompt-specific_2023} takes a step further into degrading the quality of generated output from a text-to-image model as a goal of this method is to prevent training on unauthorised data, which shall serve as a strong disincentive to disregard the copyright notices and opt-out choices. 
The authors first demonstrate that despite the massive datasets used in training diffusion models, these are vulnerable to poisoning attacks targeting specific prompts. 
Based on this observation, Nightshade uses a small number of perceptually identical poison samples to corrupt the response of a model to specific prompts. 
The protection method shows large effects on related concepts and hence destabilises general features in the model, affecting its overall image generation ability.

Anti-DreamBooth~\cite{van_le_anti-dreambooth_2023} and MetaCloak~\cite{liu_metacloak_2024} are black-box approaches based on training surrogate models to optimise the perturbations, where the objective is to force the target model to overfit on these perturbations in the fine-tuning stage. These methods are designed specifically for the unauthorised CT2I scenario.

Adversarial perturbations have shown to be effective against unauthorised instruction-guided image editing (e.g. Instruct-Pix2Pix~\cite{brooks_instructpix2pix_2023}),
by ensuring that images manipulated with instruction-guided DMs result in unrealistic outcomes~\cite{chen_editshield_2023,salman_raising_2023}. 
While copyrighted images and artworks may be subject to such edits, unauthorised editing implies a broader range of misuse, such as privacy issues and spreading misinformation.
In terms of intellectual property, these methods protect the integrity of the original work. 

Photoguard~\cite{salman_raising_2023} introduces two types of protection methods; (i) \textit{encoder attack} and (ii) \textit{diffusion attack}.
The encoder attack perturbs the input image such that the VAE of an LDM maps the image to a wrong representation in the latent space, e.g., to the representation of a random noise image. This image used to shift the representation is called a \textit{target image}. 
The generated image usually still preserves the information from the textual prompt so the approach is not as effective as a diffusion attack.
The diffusion attack aims to perturb the input image so that the \textit{final} image generated by the LDM is a specific target image (e.g., random noise), hence targeting the full diffusion process, which includes the text prompt conditioning.
Although the diffusion attack generates images that are closer to the target image, it is very computationally expensive, as the optimisation of the perturbations requires back-propagation through the full diffusion process.

EditShield~\cite{chen_editshield_2023} employs an encoder attack (c.f. Photoguard) without a specific target image. 
Instead, the approach is to maximise the distance of the perturbed representation from the original representation while constraining the amount of perturbations to keep them visually imperceptible. 
They further introduce a method for finding universal perturbations for unauthorised image editing that can be applied to protect a large number of images, without having to find optimal perturbations for each image.

\input{adversarial-perturbations}

Adversarial perturbation-based methods and their key properties are compared in \Cref{tab:adversarial-perturbations}. 
We outline the \textit{method target}, i.e. the sub-process of a generative model that the adversarial modifications affect.
The resulting \textit{distortions} in the generated content can be observed as two main types. 
Firstly, the \textit{semantic distortions} render the objective of a malicious action ineffective, while keeping the overall quality of the generated image high. 
For instance, keeping the content of an image while not adhering to textual prompts requesting a specific style of the generated image. 
Secondly, \textit{graphical distortions} compromise the perceptual quality of the generated image. 

The main challenge of protecting against style mimicry via adversarial perturbations is directing these perturbations such that they shift specifically those features related to the artistic style of the victim, i.e. achieving only semantic distortions, as the artistic style is very hard to be formally defined. 
Glaze addresses this issue by employing style transfer before computing the perturbation to target only the style features, while other methods come with the cost of degrading the quality of the generated output (quality distortion).
However, this cost can be seen as a desired consequence as it provides an additional layer of protection. 
Larger disruptions in the generated content cover the broader range of violations, specifically, they protect against the unauthorised usage of copyrighted images in the training process. 
The protection is especially effective in scenarios fine-tuning diffusion models (style transfer scenario), where the protected imagery has a more direct impact on the generated content~\cite{liang_mist_2023}.
To be effective, all methods need to be applied to images before their usage in the model, whether it is training or customisation and editing. 
This means that the responsibility of protection is put on content owners -- they need to apply these methods before sharing their content.
This entails two challenges for effective and sustainable protection: (i) the level of awareness and education needs to be significantly increased to encourage the content owners (usually artists) to apply these methods before sharing their work online, and (ii) the GAI provider may circumvent the protective effects of adversarial perturbations with the release of new versions of generative models. 
These can be addressed by shifting the responsibility to other stakeholders such as hosting platforms and AI providers and their collaboration. 
As discussed in \cite{salman_raising_2023}, one instance of such collaboration may involve the AI provider offering the cloaking/immunisation services for their models and ensuring their forward compatibility with future versions.

\subsubsection{Concept removal and unlearning}\label{sec:mitigation-concept}
\begin{figure*}[ht]
    \centering
    \hspace*{\fill}
    \subfloat[\label{fig:erasing-concepts:nudity}Erasing Nudity]{%
        \includegraphics[width = .25\linewidth]{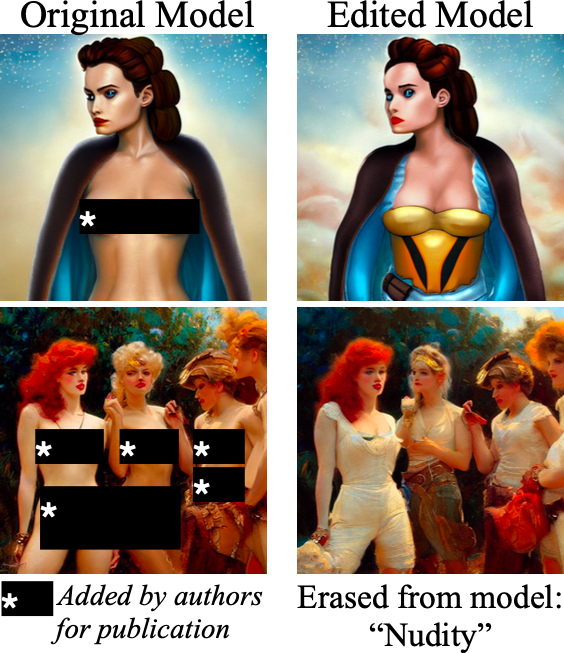}}
    \hfill
    \subfloat[\label{fig:erasing-concepts:van-gogh}Erasing Artistic Sytle]{%
        \includegraphics[width = .25\linewidth]{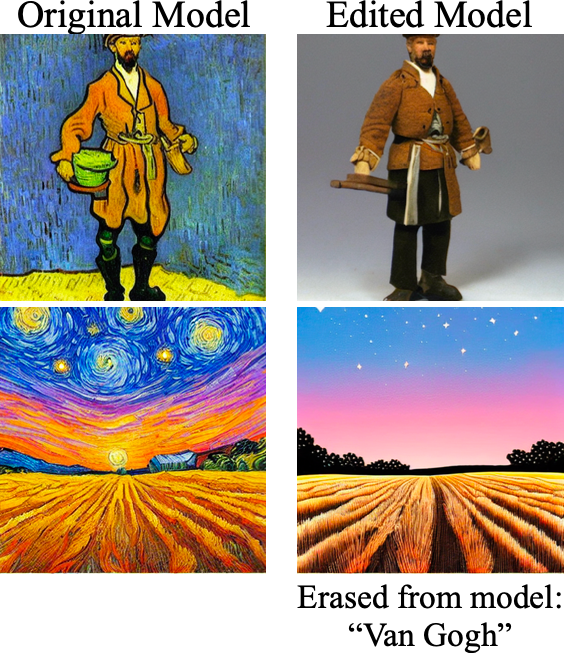}}
    \hfill
    \subfloat[\label{fig:erasing-concepts:car}Erasing Objects]{%
        \includegraphics[width = .25\linewidth]{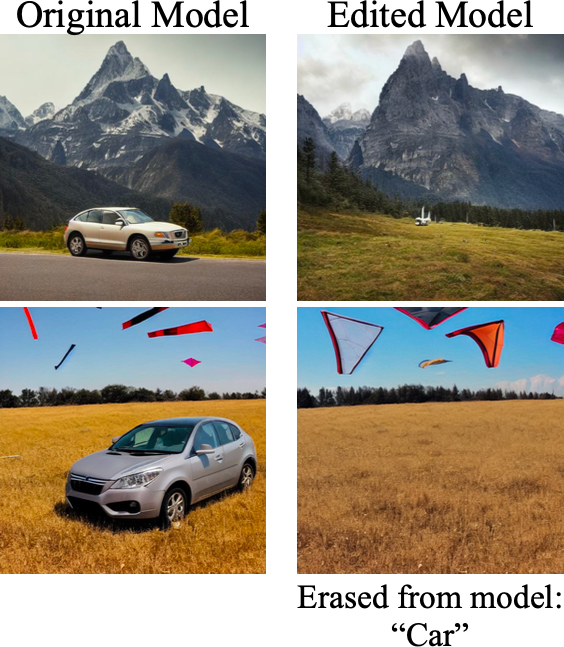}}
    \hspace*{\fill}
    
    \caption{\textbf{Concept removal}: images generated without and with \textit{erasing concepts} method~\cite{gandikota_erasing_2023}.}
    \label{fig:erasing-concepts}
\end{figure*}

\begin{figure}[ht]
    \centering
    \includegraphics[width=0.8\linewidth]{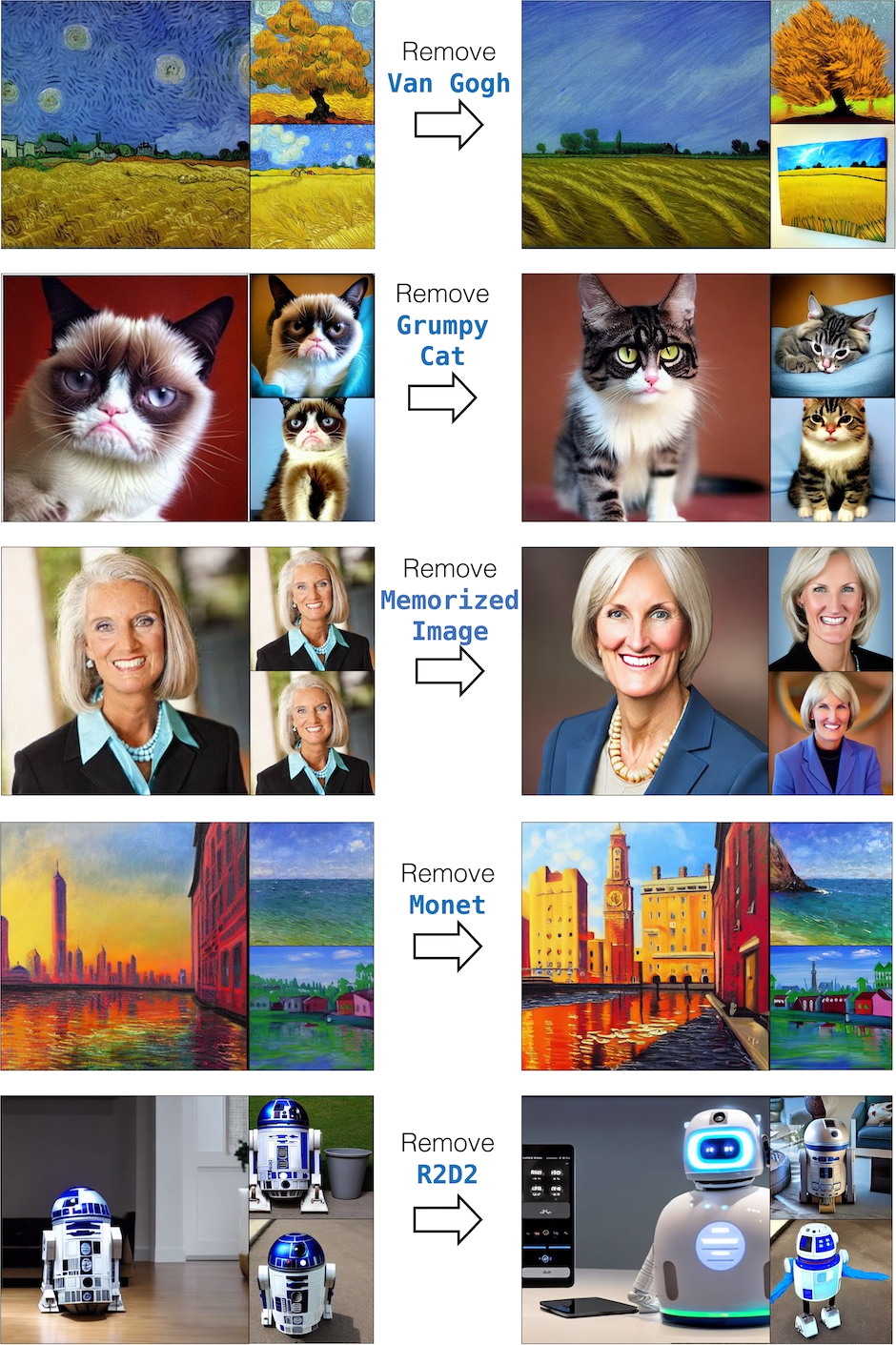}
    \caption{\textbf{Concept removal}: images generated without and with applying \textit{concept ablation}~\cite{kumari_ablating_2023}.}
    \label{fig:ablation}
\end{figure}
Earlier works that aimed to address concept removal, e.g., to counter the replication of exact objects, text embeddings, etc., in large generative models focused on the persistence of harmful elements from the training set that pass over to the generated data, such as violence and sexual content. 
Cleansing the training set from harmful samples and retraining the model from scratch is not feasible for large models already pre-trained on large data. 
Also, censoring concepts at inference time is shown to be easy to bypass~\cite{rando_red-teaming_2022}.

Current effective mitigation techniques in this category primarily focus on concept removal via fine-tuning, although some of the proposed techniques require at least a fraction of full retraining. 
For instance, the authors of the \textit{data redaction}~\cite{kong_data_2023} method aim to remove data from pre-trained GANs by re-training the network on the original set, adding the redacted data to the fake data set and applying standard adversarial loss. 

\textit{Safe latent diffusion}~\cite{schramowski_safe_2023} is proposed as a version of the Stable Diffusion (SD) model, trained to exclude inappropriate imagery originally contained in the LAION dataset used to train the SD. 
It extends a latent diffusion model by incorporating a text prompt that describes inappropriate concepts such as hate, harassment, sexual content, self-harm, violence, shocking images, and illegal activity.
This addition allows the model to simultaneously guide the image generation process towards producing content that aligns with the desired prompt and steer away from generating content associated with the defined inappropriate concepts.
SLD is a seminal work for many later concept removal techniques that specifically address IP-related issues. We discuss those in the remainder of this section.

Many concept removal methods for pre-trained models use fine-tuning to remove the target concept permanently from the model by modifying its weights while minimising the effects on other concepts.
In \cite{gandikota_erasing_2023}, \textit{erasing concepts} is based on the observation that cross-attention modules serve for text conditioning in text-to-image models (as opposed to self-attention modules that activate regardless of the text conditioning), therefore by fine-tuning these parameters the connection between text prompts and the output images can be removed. 
\Cref{fig:erasing-concepts} shows how this method can be scaled to remove harmful concepts or unwanted objects and mitigate style mimicry attacks.
Unified concept editing (UCE)~\cite{gandikota_unified_2024} is a generalised method from the same group of authors, where they combine approaches to mitigate style mimicry, bias and offensive content by concept erasure, while only causing minor effects on other concepts.   

Other fine-tuning-based approaches~\cite{kumari_ablating_2023,zhang_forget-me-not_2023} propose \textit{concept ablation} resp. \textit{concept unlearning} by directing the unwanted samples towards a more general concept.
These methods require additional input from the user, namely the description of a generalised concept in relation to the target concept - the \textit{anchor concept}. 
During fine-tuning, the objective is to teach the model to generate the output as if the anchor concept was prompted instead of the target concept (e.g., "Nemo leaping out of the water" as "A fish leaping out of the water") via minimising the statistical distance between the generated image distribution of the target concept and the image distribution of the anchor concept.
The resulting ablated images shown in \Cref{fig:ablation} lose the specific features of the target concept. 
Limitations noted by the authors include the degradation in closely surrounding concepts (e.g., the ablation of the concept "Van Gogh" highly affects the concept "Monet").

The discussed concept removal techniques for image generation demonstrate an effective mitigation of style mimicry, as the artistic style could in this sense also represent a concept. 
This paves the way for developing efficient opt-out techniques for artists who wish to secure their works against mimicry. 

Unlearning techniques for LLMs include algorithms for approximating the removal of training data without retraining the model, which would be infeasible due to the prohibitive computational demand. 
Two objectives are usually addressed in LLM unlearning~\cite{liu_rethinking_2024} - \textit{model capability removal}, mostly relevant for achieving privacy and safety~\cite{jang_knowledge_2022,wu_depn_2023,lu_quark_2022}, and \textit{data influence removal}, crucial for intellectual property protection.
One such data influence removal approach demonstrates a case study using "Harry Potter" as a targeted content~\cite{eldan_whos_2023}. 
It proposes generating generic predictions to replace specific content, thereby removing the memory of the targeted information while preserving its general performance. 

\subsubsection{Watermarking \& ownership verification} \label{sec:mitigation-watermarking}
\begin{figure}
    \centering
    \includegraphics[width=\linewidth]{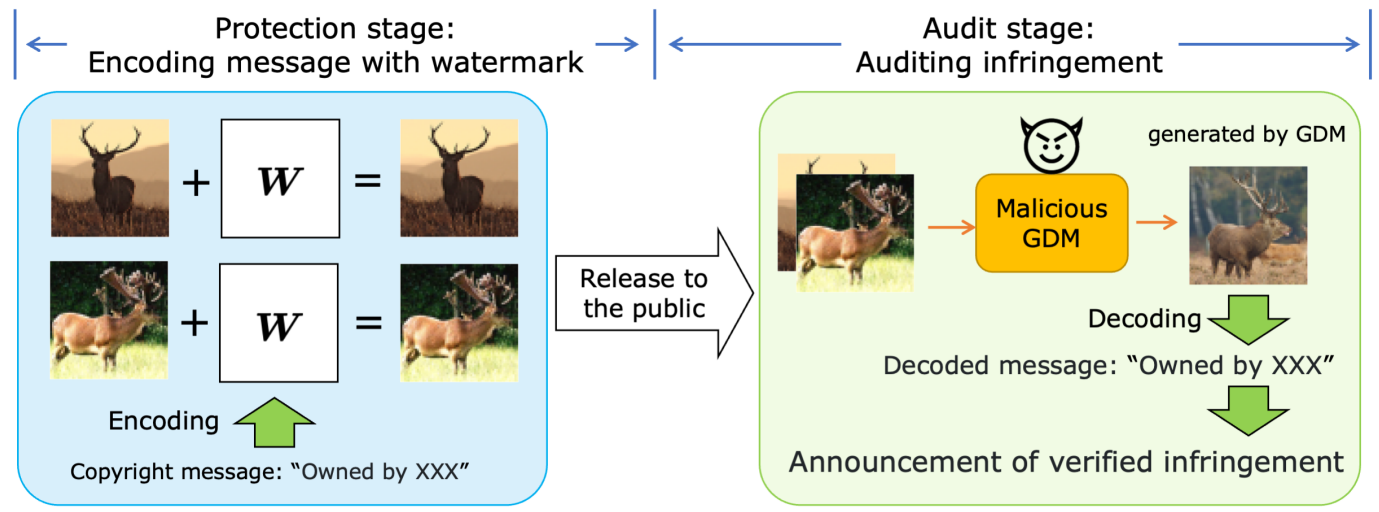}
    \caption{The two stages of forensic watermarking~\cite{cui_diffusionshield_2023}.}
    \label{fig:watermarking-stages}
\end{figure}
Watermarking is a method facilitating the detection of content that is potentially subject to unauthorised usage as training data of generative models; it involves preemptively embedding imperceptible signals or patterns into the content to assert ownership or trace unauthorised use, to eventually protect intellectual property. 
These techniques ensure that content generated by a GAI that leverages such watermarked data can be identified or verified as originating from protected sources, providing a mechanism for content owners to safeguard their work. 

Unlike methods based on adversarial perturbations, watermarking methods generally do not degrade the utility of the protected images for generation purposes but rather enable detection and verification of data ownership. 
However, there is a slight ambiguity in the terminology throughout the literature, as a few of the methods addressed in \Cref{sec:mitigation-adversarial} are referred to as watermarking as well. 
In \cite{zhang_editguard_2023} the authors refer to these methods as \textit{adversarial} watermarks, as opposed to the \textit{forensic} watermarks that are covered in this section. 
As mentioned above, forensic watermarks achieve the goals of ownership verification and traceability. 
This implies a two-stage process, as depicted in \Cref{fig:watermarking-stages}; this two-stage process is another key property of forensic watermarks that distinguishes it from adversarial watermarks.
Forensic watermarking generally consists of (i) the \textit{protection stage}, where the watermark is created and embedded into the training samples, and which needs to take place before content sharing; and (ii) the \textit{audit stage}, which includes detecting the watermark from the generated content, its verification and further actions due to the committed violations. 

Forensic watermarking has the following main requirements (analogous to machine-learning model watermarking \cite{lederer_identifying_2023}):
\begin{itemize}
    \item \textit{robustness}: the watermark is robust to changes in the content caused by the generation process or a text-guidance
    \item \textit{fidelity} (also known as \textit{invisibility}): the watermark causes little impact on the quality of both the input (protected) samples and generated (synthesised) output
    \item \textit{effectiveness}: successful watermark extraction from the watermarked content
    \item \textit{integrity}: a valid watermark cannot be extracted from a non-watermarked content
\end{itemize}

Watermarking for GANs by Hayes et al.~\cite{hayes_generating_2017} for embedding a secret signal within an image is based on adversarial training of three neural networks, \textit{A}, \textit{B} and \textit{C}. 
\textit{A} embeds the signal within the original unmodified, \textit{cover} image, producing a \textit{steganographic} image, and sends it to \textit{B}, who can recover the message. \textit{C} can eavesdrop on the link between \textit{A} and \textit{B} and aims to discover if there is a secret signal embedded within their communication; given an image, \textit{C} places a confidence score of how likely the image is cover or steganographic. 
Given this scenario, network \textit{A} is learning to embed a signal such that network \textit{B} can recover it and network \textit{c} can do no better than randomly guessing if an image is a cover or steganographic. 
The GAN watermarking techniques require full access to the training process and perform no better than random guessing for subject-driven synthesis via diffusion models~\cite{ma_generative_2023}. 

For diffusion models, GenWatermark~\cite{ma_generative_2023} is a method for protecting image data in a subject-driven synthesis scenario. 
It consists of a generator and a detector, both pre-trained on a large general dataset and optimised in an adversarial learning manner under the bound of perturbation budget; the generator is learned to generate strong watermarks while the detector learns to distinguish between clean and watermarked images. 
This concludes the first phase, after which, a generator and a generic detector have been trained. 
In the second phase, the detector can be fine-tuned to a specific subject (e.g., one artistic style). 
This is done such that two models are trained, a clean model on a clean image set and a watermarked model on a corresponding watermarked image set. 
These two models are then used to synthesise image sets for fine-tuning the detector.

Similarly, concept watermarking~\cite{feng_catch_2023} is proposed with the idea of jointly training a watermark encoder and watermark decoder to embed a watermark into a specific concept. 
This method is evaluated specifically for a textual inversion scenario, i.e. subject-driven fine-tuning, hence the watermark is directed towards the introduced word.   

FT-Shield~\cite{cui_ft-shield_2023} is a method designed for fine-tuning text-to-image diffusion models. 
The perturbations that represent a watermark are optimised such that they are learned by the diffusion model before other features such as style.
Watermark detection is done by a binary classifier learned to distinguish clean from watermarked images using the original pairs of clean and protected images and augmented with images generated by the model. 

DiffusionShield~\cite{cui_diffusionshield_2023} is a method that embeds watermarks into images and ensures that the watermark is learned by a diffusion model and reproduced in generated images, allowing the detection of a legitimate owner.
The watermark is thus able to persist the diffusion process, unlike prior image watermarking approaches based on the frequency domain~\cite{navas_dwt-dct-svd_2008} and NNs~\cite{zhu_hidden_2018}, and those designed for DeepFake detection~\cite{yu_artificial_2021}.
The authors assume a scenario where the content owner has rights over multiple images and embeds the same watermark into each of them. 
The effectiveness of the watermarks is also demonstrated in scenarios where multiple owners embed separate (different) watermarks into their respective images. 

EditGuard~\cite{zhang_editguard_2023} is a framework meant to achieve two goals simultaneously; copyright protection and tamper localisation. 
It leverages image-into-image steganography (an image embedded within another image) for embedding and accurately decoding both tampered areas and copyright information. 
Robust Invisible Watermarking (RIW)~\cite{tan_somewhat_2023} is a watermarking technique specifically designed against unauthorised editing. 
These watermarks are typically simple, repetitive images such as plain text with a black background, and are embedded into the latent representation of the target image. 
This method optimises the watermark separately for each input sample. 
After editing, the watermark can be extracted by calculating a pixel-wise difference between the watermark and the edited image or utilising an OCR (Optical Character Recognition) model, the latter being more effective.

TimbreWatermarking~\cite{liu_detecting_2024} is a method for watermarking against unauthorised audio synthesis. 
It embeds a watermark in the frequency domain that aims to be detectable after voice cloning attempts.

\subsubsection{Detection and attribution}\label{sec:mitigation-detection}

To provide the artists with an efficient and rather simple way to make initial steps in protecting their art, efforts have been put into developing tools where artists can verify whether their art is part of large datasets and, consequently, a part of training sets for generative models, i.e. the \textit{detection of participation}. 
One such solution is the project \textit{Have I been trained?}\footnote{\url{https://haveibeentrained.com/}} by the company \textit{Spawning}\footnote{\url{https://www.spawning.ai}}. 
The tool relies on Contrastive Language-Image Pre-Training (CLIP) retrieval to enable artists to search the LAION-5B and LAION-400M datasets, which are used to train Stable Diffusion and Midjourney models. 
In the Spawning framework, this detection serves as an initial step, which can be followed by an opt-out mechanism for artists to retract their art from the datasets. 

\textit{Data attribution} in generative models involves identifying the contribution of specific data samples to the generated outputs. 
From an ethical point of view, this process is crucial for understanding how models generate content, addressing not only concerns about copyright but also data privacy and transparency. 
Techniques for data attribution may include proactive detection methods such as embedding watermarks in the training data which can later be detected in the generated content, i.e. forensic watermarks discussed in \Cref{sec:mitigation-watermarking}. 
On the other hand, different analytical methods may be used to trace the influence of training samples on the behaviour of the model. 

Given a generated sample, finding the contributing matches across the large training dataset is often infeasible, however, in some special cases, the approximate matches can be extracted.
This special case requires generated content to be highly, almost 100\% influenced by a single training sample and is shown to be a feasible scenario~\cite{carlini_extracting_2023,somepalli_diffusion_2022}. 
In \Cref{sec:IPR-threats} we discussed this as a data replication problem that may be leveraged for copyright infringements. 
In this context, it can serve as evidence to the copyright owner that their data has been used for training. 

\begin{figure*}
    \centering
    \includegraphics[width=.9\linewidth]{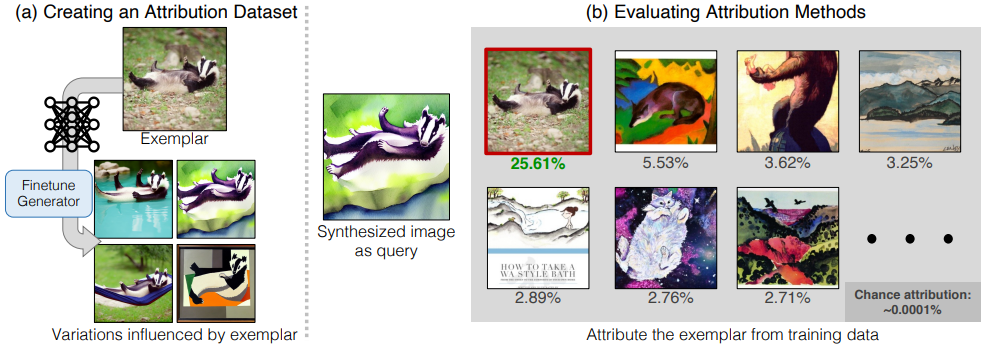}
    \caption{Attribution by Customisation (AbC)~\cite{wang_evaluating_2023}.}
    \label{fig:attribution}
\end{figure*}

When it comes to attribution in the more general but complex scenario where the generated sample is influenced by multiple training samples, Wang et al.~\cite{wang_evaluating_2023} take a leap towards understanding the influences of an individual image in the generative process. 
Drawing from the observations that the main challenge in attributing generated images to specific training samples is the lack of ground truth for this task, the key idea is to simulate ground truth directly by adding exemplar influences, as illustrated in \Cref{fig:attribution}.
Wang et al. use textual inversion to customise images, thus creating a set of synthesised images influenced by their corresponding exemplar image. 
Using the dataset, the method tunes the retrieval feature spaces CLIP~\cite{pizzi_self-supervised_2022}, DINO~\cite{caron_emerging_2021}, ViT~\cite{dosovitskiy_image_2021}, ALADIN~\cite{ruta_aladin_2021}, SSCD~\cite{pizzi_self-supervised_2022} via contrastive learning~\cite{oord_representation_2019} to obtain attribution scores representing the influence of multiple candidate images on the generated image.
In \cite{dai_training_2023} the authors introduce the \textit{encoded ensemble} method that enables the identification of influential training examples by leveraging ensemble training on carefully engineered splits of the training data. 
The main limitation of this method is its computational expansiveness as it requires training ensembles of 10-50 diffusion models.

\subsection{Policy, ethical usage and recommendations}\label{sec:mitigation-ethical}
Several sophisticated models are already trained on a large corpus of data scraped from different sources, such as portfolio sites for artists (Pinterest, ArtStation) and shopping sites (Fine Art America), without the artists' consent for their art being used as training data.

As discussed in \Cref{sec:mitigation-detection}, Spawning automates the process of identifying and flagging non-consenting data, creating opt-out lists that are forwarded to LAION who have agreed to remove those images from its dataset~\cite{wu_how_2023}.
Opt-out options have recently been adopted by online artist communities such as \textit{Deviant Art}. 
With this, they attempt to disallow third parties to scrape the content for AI development purposes without artists' knowledge and permission~\cite{wiggers_deviantart_2022}. 
In fact, the default setting for artists is an \textit{opt-out}; therefore an artist needs to make an active choice and consent to opt in. 
Some platforms even prevent the scrapping completely; The Guardian, New York Times, CNN, Australian Broadcasting Corporation, British Broadcasting Corporation and Reuters have reportedly blocked ChatGPT (GPTBot) and Common Crawl from scraping their content due to the intellectual property considerations of the authors and licence fee payers~\cite{bogle_new_2023,marcus_generative_2024}.
Some projects address the development of ethical tools for generating art content. 
These AI tools are trained explicitly on data for which consent was given.
Spawning offers an API for training ethical models ensuring that all underlying data has been provided with consent.
One instance of a consent-respecting AI model is Holly+\footnote{\href{https://holly.mirror.xyz/54ds2IiOnvthjGFkokFCoaI4EabytH9xjAYy1irHy94}{https://holly.mirror.xyz/}}, a tool which allows uploading a polyphonic audio track and generating a version of it sung by a \textit{deepfake} of its creator, Holly Herndon.

Policy development plays a critical role in navigating the development and usage of generative AI. 
Technical solutions discussed in the previous sections can be incorporated within the API of the generative models, as Salman et al.~\cite{salman_raising_2023} briefly discuss this possibility in case of cloaking the copyrighted works. 
To address more general ethical considerations surrounding GAI, Knott et al.~\cite{knott_generative_2023} advocate for a collaborative effort among policy-makers, researchers and AI developers to demonstrate and integrate reliable detection mechanisms for the content their models generate, relying on various technical approaches such as watermarking, exploiting statistical features and maintaining logs of generated content for comparison. These detection tools should be publicly available, enabling users to verify the origin of content. 
Furthermore, the \textit{Content Authenticity Initiative}\footnote{\url{https://contentauthenticity.org}} is developing an open standard and provenance tool that would prove the authenticity of digital content via watermarking. 
It promotes trustworthiness and transparency by helping fight disinformation as well as protecting the IP rights of digital content creators.

A responsible act from the perspective of model development is taking into account the concerns of memorisation and assessing the necessity of further actions (e.g., applying techniques from \Cref{sec:mitigation}) before deploying the model.
It is generally a good recommendation to test the model's generalisation ability, however, this is not a good enough indicator of whether the model still memorises the training data. 
Carlini et al. \cite{carlini_secret_2019} propose a strategy for testing the model memorisation already at training time. 
This can be done by augmenting the training process with some random unexpected input and measuring the likelihood of its generation. 
Although this method has been empirically evaluated on large text models (e.g. Google's Smart Compose, a commercial text-completion model~\cite{chen_gmail_2019}) to measure the model's potential unintended memorisation of unique sequences, it could be repurposed for copyright protection too. 
Ring-A-Bell~\cite{tsai_ring--bell_2023} is a red-teaming framework proposed for prompt-based concept testing that generates prompts leading to the generation of problematic concepts.

Formalisation of the problem of intellectual property protection in AI is addressed in the recent work ~\cite{vyas_provable_2023}. 
The authors draw parallels from differential privacy and provide a formal guarantee on the (lack of) similarity between the output of a generative model and any potentially copyrighted data in its training set. 
The method to achieve the guarantee is evaluated on a transformer and diffusion model although it is claimed as model-agnostic.

\section{Discussion}\label{sec:discussion}
The rapid development of generative AI sparked debates on several other ethical aspects of the usage of GAI beyond the discussion regarding the intellectual property of the training data; these include the generation of harmful content, privacy, AI-generated content detection, and the IP of the generated content. 
Unfiltered training data and manipulative intents can lead to violations in these areas; many protection methods address some of these issues simultaneously. 
In this section, we first discuss works addressing other ethical issues in GAI that largely overlap with IP protection and the nature of their relation. 
Secondly, we discuss the general challenges of the proposed methods for IP protection we presented in \Cref{sec:mitigation}.

\subsection{IP protection vs. other ethical questions in GAI}
Concept removal~\cite{liu_geom-erasing_2023,ho_classifier-free_2022} and unlearning~\cite{wu_erasediff_2024} are examples of the methods addressing harmful content generation -- in fact, they were originally designed for this purpose, and later also evaluated on IP protection scenarios.
Hence, we discussed concept removal methods in the light of protection against style mimicry in \Cref{sec:mitigation-concept}. 

Methods initially designed for harmful content protection that have not yet been evaluated for IP protection can likely be utilised in this scenario as well, similarly to the examples from \Cref{sec:mitigation-concept}, by treating the artistic style as a concept.
Furthermore, harmful content is addressed via detection~\cite{rando_red-teaming_2022} and prompt-modifying methods~\cite{hanu_detoxify_2020} based on safety checkers filtering the harmful content. 
These could also be extended to IP protection scenarios by detecting and filtering keywords that lead to potential copyright violations (e.g. names of the artists)~\cite{noauthor_openai_2023}.

Similarly, there are notable cloaking methods~\cite{wu_towards_2023} designed for protecting privacy against human face generation via DMs that do not explicitly evaluate their utility for the IP protection scenario, but are a valuable motivation for the methods in \Cref{sec:mitigation-adversarial}.

Protection against deepfake generation shares similarities to the works in IP protection. 
They use the same underlying methods (adversarial perturbations, watermarking), but address different scenarios.
For instance, methods that proactively counter deepfakes majorly rely on adversarial perturbations, similar to cloaking methods. 
These perturbations are optimised to disrupt the deepfake generation by causing the low quality of generated content or nullifying the functionality~\cite{ruiz_disrupting_2020,dong_restricted_2023,ruiz_practical_2023}. 
On the other hand, deepfake watermarking methods do not prevent, but rather detect the deepfakes~\cite{wang_faketagger_2021,wu_sepmark_2023}. 
This scenario can be compared with watermarking for IP protection~\cite{lederer_identifying_2023} as they share the objective of embedding a verifiable watermark in the generated content.

Detection of AI-generated content (not only deepfakes), e.g. text~\cite{tang_science_2024} and images~\cite{liu_detecting_2022}, plays an indirect role in intellectual property protection as well, especially when it comes to creative works. 
Current detection techniques for AI-generated text, however, struggle with reliability due to high false negative rates~\cite{sadasivan_can_2024,weber-wulff_testing_2023}. 
Moreover, the effectiveness significantly decreases when faced with machine translation and obfuscation techniques.
Similarly, for images, reactive detection techniques relying on statistical artefacts are also not effective~\cite{corvi_detection_2023}.
For image data, preemptive watermarking techniques are designed to embed a signal into the content to later be detected as generated content~\cite{zhu_hidden_2018,zhang_udh_2020}, or further, attributed to a specific generative model\footnote{\url{https://deepmind.google/technologies/synthid/}}.
These techniques have also shown to be insufficiently robust against minor perturbations~\cite{jiang_evading_2023}.

The ownership over AI-generated content is another ethical question that is heavily discussed at the moment and there is no clear consensus on who and if anyone should own the AI-generated content. 
Large training data including valuable content such as artwork, articles etc. is a crucial part of GAI capabilities, hence one view is that the artists and authors should be compensated or credited for the usage of their content~\cite{lnu_artists_2023}. 
Being able to credit the artists sparked the research and development of attribution methods, which could provide the original content owners with the appropriate acknowledgement and/or compensation; however, to this date, these attribution methods are not advanced enough to reliably detect the exact training sample contributing to the output of complex generative AI models, as discussed in \Cref{sec:mitigation-detection}. 
As opposed to this, others suggest that transformative use of data in generative AI should classify model training as fair use~\cite{lemley_fair_2021}.   

Another view is that the user of the generative AI should be able to own the generated content in some scenarios. Automatically generated content generally cannot be protected by law in several relevant jurisdictions, however, the real interaction between the user and the AI is more complex -- the generated outputs in question are usually a result of many sequential prompts and a mix of other artistic techniques such as photography, oil painting etc. 
There are notable works produced such that the artists included prompting AI in the process of creation alongside other techniques, e.g. the prize-winning “Théâtre D’opéra Spatial” by Jason M. Allen and "A Recent Entrance to Paradise" by Stephen Thaler.
Copyright protection has consistently been denied for these works ruling that prompting alone is not enough to claim authorship, as the user has limited control over how AI systems interpret and generate the output, disregarding the complex nature of their creation~\cite{roose_ai-generated_2022,brodkin_us_2023}.
Therefore the question remains open -- how much human input is necessary for the AI user to be qualified as an author?

\subsection{Recommendations and Challenges}\label{sec:conclusion:challenges}

\textit{Dataset sanitisation} techniques are a good first-line defence due to their effectiveness in reducing model memorisation and, consequently, the replication of training data in generated content. 
\textit{Inference-time prompt modification} strategies can further reduce replication. 
\textit{Content moderation at inference} is a computationally feasible method to address filtering unwanted concepts in the generated content but requires a significant amount of labelled data.
However, none of these methods entirely circumvent the memorisation problem and do not solve more specific IP issues such as style mimicry.
For that, \textit{concept removal} methods can be applied such that specific artistic styles cannot be generated.
A big challenge of concept removal methods is that the removal of the target concept usually affects the generation of related concepts, hence the performance of the GAI model can be reduced.
The general issue of lack of definition for an "artistic style" presents itself as a challenge for the \textit{cloaking methods}, as well. 
This is why most of the methods need another target style to steer the adversarial modification. 
Cloaking methods have gained the interest of both research and content-owner communities because these are the only prevention methods that content owners can proactively apply to protect their works. 

For ownership detection, content owners can also proactively embed watermarks. 
Watermarking is shown to be effective mostly when used in fine-tuning GAI (e.g. textual inversion), hence embedded into a narrow, specific concept, or in the image editing scenario where typically most of the original image is preserved in the output. 

A challenge for techniques such as watermarking or cloaking methods is the limited cross-model transferability, i.e. methods developed for a specific type and version of the generative AI model might not be as effective for a different model. Also, protection methods themselves will become targets for specific removal attacks, and attackers can e.g. adapt from the range of methods developed to defend against adversarial examples (e.g. \cite{meng_magnet_2017}). Recent work has shown to clean several types of perturbation added to images before using them for training a generative AI model; for example, \cite{qin_destruction-restoration_2023} showed to remove perturbations from PhotoGuard~\cite{salman_raising_2023} and Anti-DreamBooth~\cite{van_le_anti-dreambooth_2023} .

This puts the content owners into an arms race with the AI providers and potentially provides a false sense of security.
Aggravating the arms race, content (even when protected with these methods) cannot be easily retracted once published online and scraped and copied by adversaries; they thus have sufficient time to develop protection removal methods.

These types of issues call for more policy-based approaches and better collaboration between content owners, researchers and AI providers. 
This would ensure a more sustainable development of AI where the accent is on promoting trust and transparency as much as supporting innovation.

\section{Conclusions}\label{sec:conclusion}
In this paper, we reviewed potential IP violations of data used to train (large) generative AI models, as well as state-of-the-art technical protection methods.
The methods considered generally achieve diverging goals in terms of protection: (i) they may limit the usability of the data samples in the training process, (ii) provide traceability and attribution of the usage of data from the generated content and (iii) protect artistic styles from being learned and generated by the model.

IP protection is only a fragment of the trustworthy and responsible development of generative AI.
We discussed how the existing solutions for other ethical issues, such as harmful content generation and deepfakes, share similar underlying techniques with IP protection but focus on different outcomes.
In an ideal scenario, these goals could be met simultaneously. Therefore, significant challenges lay ahead towards developing trustworthy and responsible generative AI. 

While the landscape of IP protection techniques in generative AI is only beginning to materialise, we noted in \Cref{sec:conclusion:challenges} that attacks that invalidate these protections have already been proposed.
Looking forward, the longevity of generative AI technologies and their fair use will be determined not just by their technological advancements but by how well they integrate ethical considerations and comply with evolving policy landscapes as they become more deeply integrated into various sectors.
Therefore, individuals and organisations involved in generative AI need to be aware of the potential violations and ensure that they have the necessary rights and permissions for the data and content used in the development and application of these models. 
While this paper contributes to the understanding of technical IP protection methods in generative AI, it also calls for ongoing improvements in protection technologies and policies emphasising transparency and fairness in generative AI development.

\balance

\bibliographystyle{ieeetr}
\bibliography{references}  

\end{document}

%% file: taxonomy.tex
\begin{figure*}[t!]
\begin{tikzpicture}
\node[inner sep=0pt] (taxonomy) at (0,0) {\includegraphics[width=\linewidth]{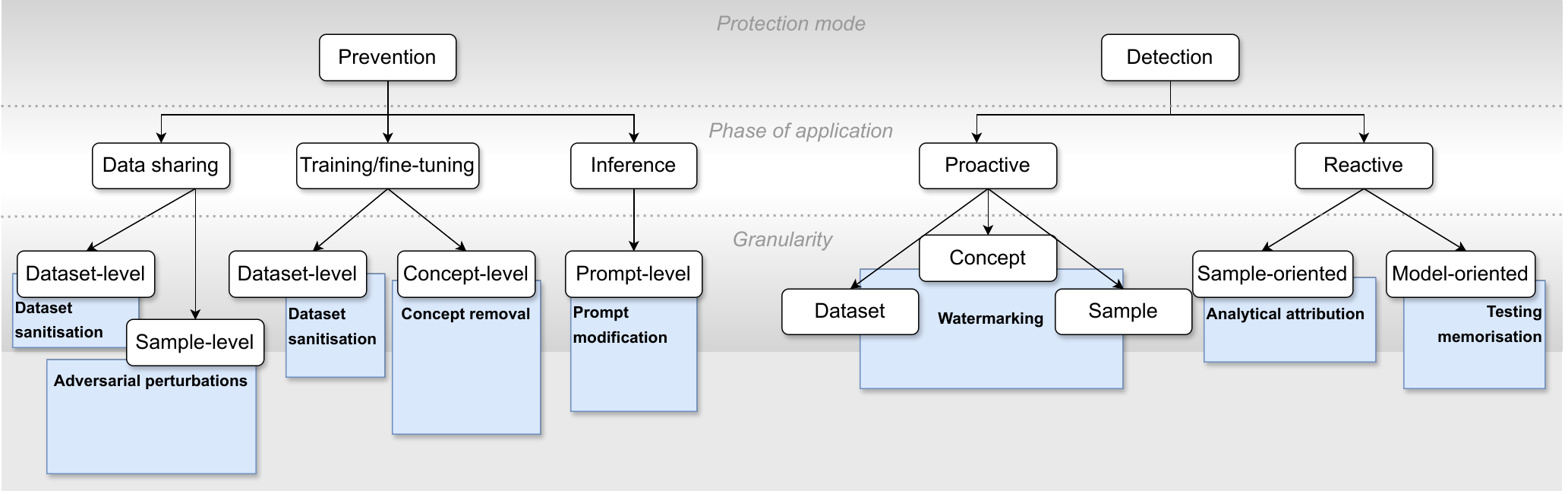}};

\node[] (sanitisation1) at (-6.76,-0.63) {\tiny \cite{carlini_extracting_2023}};

\node[] (adv1) at (-6.36,-1.60) {\tiny \cite{shan_glaze_2023,liang_adversarial_2023,liang_mist_2023,ye_duaw_2023}};
\node[] (adv2) at (-6.32,-1.85) {\tiny \cite{zhao_unlearnable_2023,chen_editshield_2023,salman_raising_2023,shan_prompt-specific_2023}};
\node[] (adv3) at (-6.56,-2.1) {\tiny \cite{zheng_understanding_2023,van_le_anti-dreambooth_2023,liu_metacloak_2024}};

\node[] (sanitisation2) at (-4.52,-1.15) {\tiny\cite{somepalli_understanding_2023,li_mitigate_2024}};

\node[] (concept1) at (-3.25,-0.95) {\tiny\cite{kong_data_2023,schramowski_safe_2023,gandikota_erasing_2023}};
\node[] (concept2) at (-3.16,-1.2){\tiny\cite{gandikota_unified_2024,kumari_ablating_2023,zhang_forget-me-not_2023}};
\node[] (concept3) at (-3.40,-1.45){\tiny\cite{eldan_whos_2023,liu_rethinking_2024}};

\node[] (prompt1) at (-1.65, -1.15) {\tiny\cite{somepalli_understanding_2023,noauthor_openai_2023}};
\node[] (prompt2) at (-1.6, -1.40) {\tiny\cite{hanu_detoxify_2020}};

\node[] (wm_dataset) at (1.0, -1.05) {\tiny\cite{cui_diffusionshield_2023}};
\node[] (wm_sample1) at (2.65, -1.05) {\tiny\cite{zhang_editguard_2023,hayes_generating_2017,ma_generative_2023}};
\node[] (wm_sample2) at (2.65, -1.30) {\tiny\cite{cui_ft-shield_2023,tan_somewhat_2023,liu_detecting_2024}};
\node[] (wm_concept) at (2.05,-0.50) {\tiny\cite{feng_catch_2023}};

\node[] (analytical_attr) at (5.03, -0.95) {\tiny\cite{somepalli_diffusion_2022,carlini_extracting_2023,wang_evaluating_2023,dai_training_2023}};

\node[] (testing) at (6.85, -1.2) {\tiny\cite{carlini_secret_2019,tsai_ring--bell_2023,vyas_provable_2023}};

\end{tikzpicture}
\caption{Taxonomy of the IP protection methods for training data in GAI.}
\label{fig:taxonomy}
\end{figure*}

%% file: terminology.tex
\begin{table*}
\centering
\caption{Overview and terminology of categories of IP protection methods.}
\label{tab:terminology}
\rowcolors{2}{white}{gray!15}
\resizebox{\textwidth}{!}{
\begin{tabular}{l|ll}
\toprule
\textbf{Category} &
  \textbf{Definition} &
  \textbf{Synonymous terms} \\ \midrule
Dataset sanitisation &
  Modifying training dataset before it is used for model training &
  - \\ \hline
Prompt modification &
  Modifying textual prompts in T2I scenario at inference time &
  - \\ \hline
Adversarial perturbations &
  \begin{tabular}[c]{@{}l@{}}Applying noise to train data samples intentionally crafted to\\ disrupt the generation process\end{tabular} &
  \begin{tabular}[c]{@{}l@{}}adversarial noise~\cite{zhao_unlearnable_2023}, style cloak / cloaking~\cite{shan_glaze_2023}, \\ poisoning~\cite{liu_metacloak_2024}, adversarial watermark(-ing)~\cite{zhang_editguard_2023}, \\ watermark(-ing)\cite{ye_duaw_2023}, immunisation~\cite{salman_raising_2023}\end{tabular} \\ \hline
Concept removal &
  \begin{tabular}[c]{@{}l@{}}Modifying the learning process to affect the downstream\\ content generation\end{tabular} &
  \begin{tabular}[c]{@{}l@{}}concept unlearning~\cite{zhang_forget-me-not_2023}, concept \\ ablation~\cite{kumari_ablating_2023}, data redaction~\cite{kong_data_2023},\\ concept erasure~\cite{gandikota_erasing_2023}\end{tabular} \\ \hline
Watermarking &
  \begin{tabular}[c]{@{}l@{}}The embedding of imperceptible signals into the content to\\ assert ownership or trace unauthorised use\end{tabular} &
  forensic watermarking~\cite{zhang_editguard_2023} \\ \hline
Analytical data attribution &
  \begin{tabular}[c]{@{}l@{}}Applying post-hoc analytical methods to identify the\\ contribution of specific train samples to the generated outputs\end{tabular} &
  - \\ \hline
Testing memorisation &
  \begin{tabular}[c]{@{}l@{}}Quantifying the memorisation capabilities of an underlying\\ GAI model\end{tabular} &
  -\\
   \bottomrule

\end{tabular}
}
\end{table*}

%% file: adversarial-perturbations.tex
\begin{table*}
    \centering
\caption{Adversarial perturbation-based protection methods and their main properties. I2I: image-to-image translation; T2I: text-to-image generation; CT2I: customised text-to-image.}
\label{tab:adversarial-perturbations}
\rowcolors{2}{white}{gray!15}
    \begin{tabular}{l|>{\raggedright\arraybackslash}p{0.19\linewidth}>{\raggedright\arraybackslash}p{0.08\linewidth}>{\raggedright\arraybackslash}p{0.23\linewidth}>{\raggedright\arraybackslash}p{0.16\linewidth}}
 \toprule
 \textbf{Method}& \textbf{Method target}& \textbf{Scenario} & \textbf{Violation} & \textbf{Distortion}\\
 \midrule
         AdvDM~\cite{liang_adversarial_2023}&  diffusion&  CT2I, I2I &  style mimicry
unauth. editing & semantic\\
         Mist~\cite{liang_mist_2023}&  VAE encoder \& diffusion&  CT2I, I2I &  style mimicry
unauth. editing& semantic/graphical \\
         UDP~\cite{zhao_unlearnable_2023}&  diffusion& T2I, CT2I &  style mimicry
unauth. training&graphical \& semantic\\
         Glaze~\cite{shan_glaze_2023}&  VAE encoder& T2I, CT2I &  style mimicry& semantic\\
         DUAW~\cite{ye_duaw_2023}&  VAE decoder& CT2I &  style mimicry
unauth. training& graphical\\
Nightshade~\cite{shan_prompt-specific_2023} & VAE encoder & T2I & unauth. training   & graphical \& semantic \\
 Anti-Dreambooth~\cite{van_le_anti-dreambooth_2023} & fine-tuning process & CT2I & unauth. training & graphical \& semantic\\ 
 Metacloak~\cite{liu_metacloak_2024} & fine-tuning process & CT2I & unauth. training & graphical \& semantic\\
 ACE~\cite{zheng_understanding_2023} & fine-tuning process & CT2I & style mimicry, unauth. training & graphical\\
 EditShield~\cite{chen_editshield_2023}&  VAE encoder&  I2I&  unauth. editing& semantic\\
 Photoguard~\cite{salman_raising_2023}& VAE encoder \& diffusion& I2I& unauth. editing& graphical\\
 \bottomrule
    \end{tabular}

\end{table*}

%% file: main.bbl
\begin{thebibliography}{100}

\bibitem{openai_gpt-4_2023}
OpenAI, ``{GPT}-4 {Technical} {Report},'' Mar. 2023.
\newblock \href{http://arxiv.org/abs/2303.08774}{arXiv:2303.08774}.

\bibitem{rombach_high-resolution_2022}
R.~Rombach, A.~Blattmann, D.~Lorenz, P.~Esser, and B.~Ommer, ``High-{Resolution} {Image} {Synthesis} with {Latent} {Diffusion} {Models},'' in {\em {IEEE}/{CVF} {Conference} on {Computer} {Vision} and {Pattern} {Recognition} ({CVPR})}, (New Orleans, LA, USA), IEEE, June 2022.

\bibitem{ramesh_hierarchical_2022}
A.~Ramesh, P.~Dhariwal, A.~Nichol, C.~Chu, and M.~Chen, ``Hierarchical {Text}-{Conditional} {Image} {Generation} with {CLIP} {Latents},'' Apr. 2022.
\newblock \href{http://arxiv.org/abs/2204.06125}{arXiv:2204.06125}.

\bibitem{heikkila_this_2022}
M.~Heikkilä, ``This artist is dominating {AI}-generated art. {And} he’s not happy about it.,'' Sept. 2022.
\newblock https://www.technologyreview.com/2022/09/16/1059598/ (accessed 2024-01-31).

\bibitem{nolan_ai_2023}
B.~Nolan, ``{AI} art generators face separate copyright lawsuits from {Getty} {Images} and a group of artists,'' Jan. 2023.
\newblock https://www.businessinsider.com/ai-art-artists-getty-images-lawsuits-stable-diffusion-2023-1 (accessed 2024-01-31).

\bibitem{matsumoto_membership_2023}
T.~Matsumoto, T.~Miura, and N.~Yanai, ``Membership {Inference} {Attacks} against {Diffusion} {Models},'' Mar. 2023.
\newblock \href{http://arxiv.org/abs/2302.03262}{arXiv:2302.03262}.

\bibitem{gandikota_erasing_2023}
R.~Gandikota, J.~Materzynska, J.~Fiotto-Kaufman, and D.~Bau, ``Erasing {Concepts} from {Diffusion} {Models},'' June 2023.
\newblock \href{http://arxiv.org/abs/2303.07345}{arXiv:2303.07345}.

\bibitem{wang_survey_2023}
Y.~Wang, Y.~Pan, M.~Yan, Z.~Su, and T.~H. Luan, ``A {Survey} on {ChatGPT}: {AI}–{Generated} {Contents}, {Challenges}, and {Solutions},'' {\em IEEE Open Journal of the Computer Society}, vol.~4, 2023.

\bibitem{chen_challenges_2023}
C.~Chen, Z.~Wu, Y.~Lai, W.~Ou, T.~Liao, and Z.~Zheng, ``Challenges and {Remedies} to {Privacy} and {Security} in {AIGC}: {Exploring} the {Potential} of {Privacy} {Computing}, {Blockchain}, and {Beyond},'' June 2023.
\newblock \href{http://arxiv.org/abs/2306.00419}{arXiv:2306.00419}.

\bibitem{chen_pathway_2023}
C.~Chen, J.~Fu, and L.~Lyu, ``A {Pathway} {Towards} {Responsible} {AI} {Generated} {Content},'' Dec. 2023.
\newblock \href{http://arxiv.org/abs/2303.01325}{arXiv:2303.01325}.

\bibitem{wang_security_2023}
T.~Wang, Y.~Zhang, S.~Qi, R.~Zhao, Z.~Xia, and J.~Weng, ``Security and {Privacy} on {Generative} {Data} in {AIGC}: {A} {Survey},'' Sept. 2023.
\newblock \href{http://arxiv.org/abs/2309.09435}{arXiv:2309.09435}.

\bibitem{zhong_copyright_2023}
H.~Zhong, J.~Chang, Z.~Yang, T.~Wu, P.~C. Mahawaga~Arachchige, C.~Pathmabandu, and M.~Xue, ``Copyright {Protection} and {Accountability} of {Generative} {AI}: {Attack}, {Watermarking} and {Attribution},'' in {\em Companion {Proceedings} of the {ACM} {Web} {Conference} 2023}, (Austin TX USA), ACM, Apr. 2023.

\bibitem{smits_generative_2022}
J.~Smits and T.~Borghuis, ``Generative {AI} and {Intellectual} {Property} {Rights},'' in {\em Law and {Artificial} {Intelligence}}, vol.~35, The Hague: T.M.C. Asser Press, 2022.

\bibitem{chesterman_good_2023}
S.~Chesterman, ``Good {Models} {Borrow}, {Great} {Models} {Steal}: {Intellectual} {Property} {Rights} and {Generative} {AI},'' {\em SSRN Electronic Journal}, 2023.

\bibitem{hristov_artificial_2019}
K.~Hristov, ``Artificial {Intelligence} and the {Copyright} {Survey},'' {\em SSRN Electronic Journal}, 2019.

\bibitem{zhang_adversarial_2019}
J.~Zhang and C.~Li, ``Adversarial {Examples}: {Opportunities} and {Challenges},'' {\em IEEE Transactions on Neural Networks and Learning Systems}, 2019.

\bibitem{cina_wild_2023}
A.~E. Cinà, K.~Grosse, A.~Demontis, S.~Vascon, W.~Zellinger, B.~A. Moser, A.~Oprea, B.~Biggio, M.~Pelillo, and F.~Roli, ``Wild {Patterns} {Reloaded}: {A} {Survey} of {Machine} {Learning} {Security} against {Training} {Data} {Poisoning},'' {\em ACM Computing Surveys}, vol.~55, Dec. 2023.

\bibitem{sun_adversarial_2023}
H.~Sun, T.~Zhu, Z.~Zhang, D.~Jin, P.~Xiong, and W.~Zhou, ``Adversarial {Attacks} {Against} {Deep} {Generative} {Models} on {Data}: {A} {Survey},'' {\em IEEE Transactions on Knowledge and Data Engineering}, vol.~35, Apr. 2023.

\bibitem{lederer_identifying_2023}
I.~Lederer, R.~Mayer, and A.~Rauber, ``Identifying {Appropriate} {Intellectual} {Property} {Protection} {Mechanisms} for {Machine} {Learning} {Models}: {A} {Systematization} of {Watermarking}, {Fingerprinting}, {Model} {Access}, and {Attacks},'' {\em IEEE Transactions on Neural Networks and Learning Systems}, 2023.

\bibitem{regazzoni_protecting_2021}
F.~Regazzoni, P.~Palmieri, F.~Smailbegovic, R.~Cammarota, and I.~Polian, ``Protecting artificial intelligence {IPs}: a survey of watermarking and fingerprinting for machine learning,'' {\em CAAI Transactions on Intelligence Technology}, vol.~6, June 2021.

\bibitem{kitchenham_guidelines_2007}
B.~Kitchenham and S.~Charters, ``Guidelines for performing {Systematic} {Literature} {Reviews} in {Software} {Engineering},'' Tech. Rep. EBSE-2007-01, Department of Computer Science, University of Durham, Durham, UK, 2007.

\bibitem{wohlin_guidelines_2014}
C.~Wohlin, ``Guidelines for snowballing in systematic literature studies and a replication in software engineering,'' in {\em 18th {International} {Conference} on {Evaluation} and {Assessment} in {Software} {Engineering}}, (London, England, United Kingdom), ACM, May 2014.

\bibitem{kingma_auto-encoding_2022}
D.~P. Kingma and M.~Welling, ``Auto-{Encoding} {Variational} {Bayes},'' Dec. 2022.
\newblock \href{http://arxiv.org/abs/1312.6114}{arXiv:1312.6114}.

\bibitem{bao_cvae-gan_2017}
J.~Bao, D.~Chen, F.~Wen, H.~Li, and G.~Hua, ``{CVAE}-{GAN}: {Fine}-{Grained} {Image} {Generation} through {Asymmetric} {Training},'' Oct. 2017.
\newblock \href{http://arxiv.org/abs/1703.10155}{arXiv:1703.10155}.

\bibitem{pandey_diffusevae_2022}
K.~Pandey, A.~Mukherjee, P.~Rai, and A.~Kumar, ``{DiffuseVAE}: {Efficient}, {Controllable} and {High}-{Fidelity} {Generation} from {Low}-{Dimensional} {Latents},'' Nov. 2022.
\newblock \href{http://arxiv.org/abs/2201.00308}{arXiv:2201.00308}.

\bibitem{goodfellow_generative_2014}
I.~Goodfellow, J.~Pouget-Abadie, M.~Mirza, B.~Xu, D.~Warde-Farley, S.~Ozair, A.~Courville, and Y.~Bengio, ``Generative {Adversarial} {Nets},'' in {\em Advances in {Neural} {Information} {Processing} {Systems}}, vol.~27, Curran Associates, Inc., 2014.

\bibitem{karras_progressive_2018}
T.~Karras, T.~Aila, S.~Laine, and J.~Lehtinen, ``Progressive {Growing} of {GANs} for {Improved} {Quality}, {Stability}, and {Variation},'' Feb. 2018.
\newblock \href{http://arxiv.org/abs/1710.10196}{arXiv:1710.10196}.

\bibitem{brock_large_2019}
A.~Brock, J.~Donahue, and K.~Simonyan, ``Large {Scale} {GAN} {Training} for {High} {Fidelity} {Natural} {Image} {Synthesis},'' Feb. 2019.
\newblock \href{http://arxiv.org/abs/1809.11096}{arXiv:1809.11096}.

\bibitem{karras_style-based_2019}
T.~Karras, S.~Laine, and T.~Aila, ``A {Style}-{Based} {Generator} {Architecture} for {Generative} {Adversarial} {Networks},'' Mar. 2019.
\newblock \href{http://arxiv.org/abs/1812.04948}{arXiv:1812.04948}.

\bibitem{zhu_unpaired_2020}
J.-Y. Zhu, T.~Park, P.~Isola, and A.~A. Efros, ``Unpaired {Image}-to-{Image} {Translation} using {Cycle}-{Consistent} {Adversarial} {Networks},'' Aug. 2020.
\newblock \href{http://arxiv.org/abs/1703.10593}{arXiv:1703.10593}.

\bibitem{choi_stargan_2018}
Y.~Choi, M.~Choi, M.~Kim, J.-W. Ha, S.~Kim, and J.~Choo, ``{StarGAN}: {Unified} {Generative} {Adversarial} {Networks} for {Multi}-{Domain} {Image}-to-{Image} {Translation},'' Sept. 2018.
\newblock \href{http://arxiv.org/abs/1711.09020}{arXiv:1711.09020}.

\bibitem{sohl-dickstein_deep_2015}
J.~Sohl-Dickstein, E.~A. Weiss, N.~Maheswaranathan, and S.~Ganguli, ``Deep {Unsupervised} {Learning} using {Nonequilibrium} {Thermodynamics},'' Nov. 2015.
\newblock \href{http://arxiv.org/abs/1503.03585 [cond-mat, q-bio, stat]}{arXiv:1503.03585 [cond-mat, q-bio, stat]}.

\bibitem{ho_denoising_2020}
J.~Ho, A.~Jain, and P.~Abbeel, ``Denoising {Diffusion} {Probabilistic} {Models},'' in {\em Advances in {Neural} {Information} {Processing} {Systems}}, vol.~33, Curran Associates, Inc., 2020.

\bibitem{nichol_improved_2021}
A.~Nichol and P.~Dhariwal, ``Improved {Denoising} {Diffusion} {Probabilistic} {Models},'' Feb. 2021.
\newblock \href{http://arxiv.org/abs/2102.09672}{arXiv:2102.09672}.

\bibitem{dhariwal_diffusion_2021}
P.~Dhariwal and A.~Nichol, ``Diffusion {Models} {Beat} {GANs} on {Image} {Synthesis},'' in {\em Advances in {Neural} {Information} {Processing} {Systems}}, vol.~34, Curran Associates, Inc., 2021.

\bibitem{radford_learning_2021}
A.~Radford, J.~W. Kim, C.~Hallacy, A.~Ramesh, G.~Goh, S.~Agarwal, G.~Sastry, A.~Askell, P.~Mishkin, J.~Clark, G.~Krueger, and I.~Sutskever, ``Learning {Transferable} {Visual} {Models} {From} {Natural} {Language} {Supervision},'' in {\em 38th {International} {Conference} on {Machine} {Learning}}, PMLR, July 2021.
\newblock ISSN: 2640-3498.

\bibitem{vaswani_attention_2017}
A.~Vaswani, N.~Shazeer, N.~Parmar, J.~Uszkoreit, L.~Jones, A.~N. Gomez, L.~Kaiser, and I.~Polosukhin, ``Attention is {All} you {Need},'' in {\em Advances in {Neural} {Information} {Processing} {Systems}}, vol.~30, Curran Associates, Inc., 2017.

\bibitem{radford_improving_nodate}
A.~Radford, K.~Narasimhan, T.~Salimans, and I.~Sutskever, ``Improving {Language} {Understanding} by {Generative} {Pre}-{Training}.''
\newblock https://openai.com/research/language-unsupervised (accessed 2021-01-31).

\bibitem{gemini_team_gemini_2023}
G.~Team, ``Gemini: {A} {Family} of {Highly} {Capable} {Multimodal} {Models},'' Dec. 2023.
\newblock \href{http://arxiv.org/abs/2312.11805}{arXiv:2312.11805}.

\bibitem{touvron_llama_2023}
H.~Touvron, T.~Lavril, G.~Izacard, X.~Martinet, M.-A. Lachaux, T.~Lacroix, B.~Rozière, N.~Goyal, E.~Hambro, F.~Azhar, A.~Rodriguez, A.~Joulin, E.~Grave, and G.~Lample, ``{LLaMA}: {Open} and {Efficient} {Foundation} {Language} {Models},'' Feb. 2023.
\newblock \href{http://arxiv.org/abs/2302.13971}{arXiv:2302.13971}.

\bibitem{losio_first_2022}
R.~Losio, ``First {Open} {Source} {Copyright} {Lawsuit} {Chal}­lenges {GitHub} {Copi}­lot,'' Nov. 2022.
\newblock https://www.infoq.com/news/2022/11/lawsuit-github-copilot/ (accessed 2021-01-31).

\bibitem{brooks_instructpix2pix_2023}
T.~Brooks, A.~Holynski, and A.~A. Efros, ``{InstructPix2Pix}: {Learning} to {Follow} {Image} {Editing} {Instructions},'' in {\em {IEEE}/{CVF} {Conference} on {Computer} {Vision} and {Pattern} {Recognition} ({CVPR})}, (Vancouver, BC, Canada), IEEE, June 2023.

\bibitem{zhang_hive_2023}
S.~Zhang, X.~Yang, Y.~Feng, C.~Qin, C.-C. Chen, N.~Yu, Z.~Chen, H.~Wang, S.~Savarese, S.~Ermon, C.~Xiong, and R.~Xu, ``{HIVE}: {Harnessing} {Human} {Feedback} for {Instructional} {Visual} {Editing},'' Mar. 2023.
\newblock \href{http://arxiv.org/abs/2303.09618}{arXiv:2303.09618}.

\bibitem{avrahami_blended_2022}
O.~Avrahami, D.~Lischinski, and O.~Fried, ``Blended {Diffusion} for {Text}-driven {Editing} of {Natural} {Images},'' in {\em {IEEE}/{CVF} {Conference} on {Computer} {Vision} and {Pattern} {Recognition} ({CVPR})}, (New Orleans, LA, USA), IEEE, June 2022.

\bibitem{kawar_imagic_2023}
B.~Kawar, S.~Zada, O.~Lang, O.~Tov, H.~Chang, T.~Dekel, I.~Mosseri, and M.~Irani, ``Imagic: {Text}-{Based} {Real} {Image} {Editing} with {Diffusion} {Models},'' in {\em 2023 {IEEE}/{CVF} {Conference} on {Computer} {Vision} and {Pattern} {Recognition} ({CVPR})}, (Vancouver, BC, Canada), IEEE, June 2023.

\bibitem{grynbaum_new_2023}
M.~M. Grynbaum and R.~Mac, ``New {York} {Times} {Sues} {OpenAI} and {Microsoft} {Over} {Use} of {Copyrighted} {Work},'' Dec. 2023.
\newblock https://www.nytimes.com/2023/12/27/business/media/new-york-times-open-ai-microsoft-lawsuit.html (accessed 2024-01-31).

\bibitem{brittain_judge_2023}
B.~Brittain, ``Judge pares down artists' {AI} copyright lawsuit against {Midjourney}, {Stability} {AI},'' {\em Reuters}, Oct. 2023.
\newblock https://www.reuters.com/legal/litigation/judge-pares-down-artists-ai-copyright-lawsuit-against-midjourney-stability-ai-2023-10-30/ (accessed 2024-01-31).

\bibitem{gal_image_2022}
R.~Gal, Y.~Alaluf, Y.~Atzmon, O.~Patashnik, A.~H. Bermano, G.~Chechik, and D.~Cohen-Or, ``An {Image} is {Worth} {One} {Word}: {Personalizing} {Text}-to-{Image} {Generation} using {Textual} {Inversion},'' Aug. 2022.
\newblock \href{http://arxiv.org/abs/2208.01618}{arXiv:2208.01618}.

\bibitem{lee_language_2023}
J.~Lee, T.~Le, J.~Chen, and D.~Lee, ``Do {Language} {Models} {Plagiarize}?,'' in {\em Proceedings of the {ACM} {Web} {Conference} 2023}, (Austin TX USA), ACM, Apr. 2023.

\bibitem{qian_autovc_2019}
K.~Qian, Y.~Zhang, S.~Chang, X.~Yang, and M.~Hasegawa-Johnson, ``{AutoVC}: {Zero}-{Shot} {Voice} {Style} {Transfer} with {Only} {Autoencoder} {Loss},'' in {\em 36th {International} {Conference} on {Machine} {Learning}}, PMLR, May 2019.
\newblock ISSN: 2640-3498.

\bibitem{baio_invasive_2022}
A.~Baio, ``Invasive {Diffusion}: {How} one unwilling illustrator found herself turned into an {AI} model,'' Nov. 2022.
\newblock https://waxy.org/2022/11/invasive-diffusion-how-one-unwilling-illustrator-found-herself-turned-into-an-ai-model/ (accessed 2024-01-31).

\bibitem{ruiz_dreambooth_2023}
N.~Ruiz, Y.~Li, V.~Jampani, Y.~Pritch, M.~Rubinstein, and K.~Aberman, ``{DreamBooth}: {Fine} {Tuning} {Text}-to-{Image} {Diffusion} {Models} for {Subject}-{Driven} {Generation},'' in {\em {IEEE}/{CVF} {Conference} on {Computer} {Vision} and {Pattern} {Recognition} ({CVPR})}, (Vancouver, BC, Canada), IEEE, June 2023.

\bibitem{hu_lora_2021}
E.~J. Hu, Y.~Shen, P.~Wallis, Z.~Allen-Zhu, Y.~Li, S.~Wang, L.~Wang, and W.~Chen, ``{LoRA}: {Low}-{Rank} {Adaptation} of {Large} {Language} {Models},'' Oct. 2021.
\newblock \href{http://arxiv.org/abs/2106.09685}{arXiv:2106.09685}.

\bibitem{carlini_extracting_2021}
N.~Carlini, F.~Tramèr, E.~Wallace, M.~Jagielski, A.~Herbert-Voss, K.~Lee, A.~Roberts, T.~Brown, D.~Song, U.~Erlingsson, A.~Oprea, and C.~Raffel, ``Extracting {Training} {Data} from {Large} {Language} {Models},'' in {\em 30th {USENIX} {Security} {Symposium}}, USENIX Association, Aug. 2021.

\bibitem{mccoy_how_2023}
R.~T. McCoy, P.~Smolensky, T.~Linzen, J.~Gao, and A.~Celikyilmaz, ``How {Much} {Do} {Language} {Models} {Copy} {From} {Their} {Training} {Data}? {Evaluating} {Linguistic} {Novelty} in {Text} {Generation} {Using} {RAVEN},'' {\em Transactions of the Association for Computational Linguistics}, vol.~11, June 2023.

\bibitem{kaneko_cyclegan-vc2_2019}
T.~Kaneko, H.~Kameoka, K.~Tanaka, and N.~Hojo, ``Cyclegan-{VC2}: {Improved} {Cyclegan}-based {Non}-parallel {Voice} {Conversion},'' in {\em {IEEE} {International} {Conference} on {Acoustics}, {Speech} and {Signal} {Processing} ({ICASSP})}, (Brighton, United Kingdom), IEEE, May 2019.

\bibitem{wang_tacotron_2017}
Y.~Wang, R.~J. Skerry-Ryan, D.~Stanton, Y.~Wu, R.~J. Weiss, N.~Jaitly, Z.~Yang, Y.~Xiao, Z.~Chen, S.~Bengio, Q.~Le, Y.~Agiomyrgiannakis, R.~Clark, and R.~A. Saurous, ``Tacotron: {Towards} {End}-to-{End} {Speech} {Synthesis},'' Apr. 2017.
\newblock \href{http://arxiv.org/abs/1703.10135}{arXiv:1703.10135}.

\bibitem{ren_fastspeech_2022}
Y.~Ren, C.~Hu, X.~Tan, T.~Qin, S.~Zhao, Z.~Zhao, and T.-Y. Liu, ``{FastSpeech} 2: {Fast} and {High}-{Quality} {End}-to-{End} {Text} to {Speech},'' Aug. 2022.
\newblock \href{http://arxiv.org/abs/2006.04558}{arXiv:2006.04558}.

\bibitem{brewster_fraudsters_2021}
T.~Brewster, ``Fraudsters {Cloned} {Company} {Director}’s {Voice} {In} \$35 {Million} {Heist}, {Police} {Find},'' Oct. 2021.
\newblock https://www.forbes.com/sites/thomasbrewster/2021/10/14/huge-bank-fraud-uses-deep-fake-voice-tech-to-steal-millions/ (accessed 2024-01-31).

\bibitem{somepalli_understanding_2023}
G.~Somepalli, V.~Singla, M.~Goldblum, J.~Geiping, and T.~Goldstein, ``Understanding and {Mitigating} {Copying} in {Diffusion} {Models},'' May 2023.
\newblock \href{http://arxiv.org/abs/2305.20086}{arXiv:2305.20086}.

\bibitem{somepalli_diffusion_2022}
G.~Somepalli, V.~Singla, M.~Goldblum, J.~Geiping, and T.~Goldstein, ``Diffusion {Art} or {Digital} {Forgery}? {Investigating} {Data} {Replication} in {Diffusion} {Models},'' Dec. 2022.
\newblock \href{http://arxiv.org/abs/2212.03860}{arXiv:2212.03860}.

\bibitem{carlini_extracting_2023}
N.~Carlini, J.~Hayes, M.~Nasr, M.~Jagielski, V.~Sehwag, F.~Tramèr, B.~Balle, D.~Ippolito, and E.~Wallace, ``Extracting {Training} {Data} from {Diffusion} {Models},'' Jan. 2023.
\newblock \href{http://arxiv.org/abs/2301.13188}{arXiv:2301.13188}.

\bibitem{ulku_kahveci_attribution_2023}
Z.~Ülkü Kahveci, ``Attribution problem of generative {AI}: a view from {US} copyright law,'' {\em Journal of Intellectual Property Law \& Practice}, vol.~18, no.~11, 2023.

\bibitem{vincent_getty_2023}
J.~Vincent, ``Getty {Images} sues {AI} art generator {Stable} {Diffusion} in the {US} for copyright infringement,'' Feb. 2023.
\newblock https://www.theverge.com/2023/2/6/23587393/ai-art-copyright-lawsuit-getty-images-stable-diffusion (accessed 2024-01-31).

\bibitem{feldman_does_2020}
V.~Feldman, ``Does learning require memorization? a short tale about a long tail,'' in {\em 52nd {Annual} {ACM} {SIGACT} {Symposium} on {Theory} of {Computing}}, (Chicago IL USA), ACM, June 2020.

\bibitem{feldman_what_2020}
V.~Feldman and C.~Zhang, ``What {Neural} {Networks} {Memorize} and {Why}: {Discovering} the {Long} {Tail} via {Influence} {Estimation},'' in {\em Advances in {Neural} {Information} {Processing} {Systems}}, vol.~33, Curran Associates, Inc., 2020.

\bibitem{arpit_closer_2017}
D.~Arpit, S.~Jastrzębski, N.~Ballas, D.~Krueger, E.~Bengio, M.~S. Kanwal, T.~Maharaj, A.~Fischer, A.~Courville, Y.~Bengio, and S.~Lacoste-Julien, ``A {Closer} {Look} at {Memorization} in {Deep} {Networks},'' in {\em 34th {International} {Conference} on {Machine} {Learning}}, vol.~70 of {\em Proceedings of {Machine} {Learning} {Research}}, PMLR, Aug. 2017.

\bibitem{van_den_burg_memorization_2021}
G.~van~den Burg and C.~Williams, ``On {Memorization} in {Probabilistic} {Deep} {Generative} {Models},'' in {\em Advances in {Neural} {Information} {Processing} {Systems}}, vol.~34, Curran Associates, Inc., 2021.

\bibitem{tirumala_memorization_2022}
K.~Tirumala, A.~Markosyan, L.~Zettlemoyer, and A.~Aghajanyan, ``Memorization {Without} {Overfitting}: {Analyzing} the {Training} {Dynamics} of {Large} {Language} {Models},'' in {\em Advances in {Neural} {Information} {Processing} {Systems}}, vol.~35, Curran Associates, Inc., 2022.

\bibitem{webster_this_2021}
R.~Webster, J.~Rabin, L.~Simon, and F.~Jurie, ``This {Person} ({Probably}) {Exists}. {Identity} {Membership} {Attacks} {Against} {GAN} {Generated} {Faces},'' July 2021.
\newblock \href{http://arxiv.org/abs/2107.06018}{arXiv:2107.06018}.

\bibitem{jagielski_measuring_2023}
M.~Jagielski, O.~Thakkar, F.~Tramèr, D.~Ippolito, K.~Lee, N.~Carlini, E.~Wallace, S.~Song, A.~Thakurta, N.~Papernot, and C.~Zhang, ``Measuring {Forgetting} of {Memorized} {Training} {Examples},'' May 2023.
\newblock \href{http://arxiv.org/abs/2207.00099}{arXiv:2207.00099}.

\bibitem{carlini_quantifying_2023}
N.~Carlini, D.~Ippolito, M.~Jagielski, K.~Lee, F.~Tramer, and C.~Zhang, ``Quantifying {Memorization} {Across} {Neural} {Language} {Models},'' Mar. 2023.
\newblock \href{http://arxiv.org/abs/2202.07646}{arXiv:2202.07646}.

\bibitem{yoon_diffusion_2023}
T.~Yoon, J.~Y. Choi, S.~Kwon, and E.~K. Ryu, ``Diffusion {Probabilistic} {Models} {Generalize} when {They} {Fail} to {Memorize},'' in {\em {ICML} 2023 {Workshop} on {Structured} {Probabilistic} {Inference} \& {Generative} {Modeling}}, (Honolulu, HI, USA), July 2023.

\bibitem{gu_memorization_2023}
X.~Gu, C.~Du, T.~Pang, C.~Li, M.~Lin, and Y.~Wang, ``On {Memorization} in {Diffusion} {Models},'' Oct. 2023.
\newblock \href{http://arxiv.org/abs/2310.02664}{arXiv:2310.02664}.

\bibitem{pizzi_self-supervised_2022}
E.~Pizzi, S.~D. Roy, S.~N. Ravindra, P.~Goyal, and M.~Douze, ``A {Self}-{Supervised} {Descriptor} for {Image} {Copy} {Detection},'' in {\em {IEEE}/{CVF} {Conference} on {Computer} {Vision} and {Pattern} {Recognition} ({CVPR})}, (New Orleans, LA, USA), IEEE, June 2022.

\bibitem{wen_canary_2023}
Y.~Wen, A.~Bansal, H.~Kazemi, E.~Borgnia, M.~Goldblum, J.~Geiping, and T.~Goldstein, ``Canary in a {Coalmine}: {Better} {Membership} {Inference} with {Ensembled} {Adversarial} {Queries},'' June 2023.
\newblock \href{http://arxiv.org/abs/2210.10750}{arXiv:2210.10750}.

\bibitem{akbar_beware_2023}
M.~U. Akbar, W.~Wang, and A.~Eklund, ``Beware of diffusion models for synthesizing medical images -- {A} comparison with {GANs} in terms of memorizing brain {MRI} and chest x-ray images,'' Oct. 2023.
\newblock \href{http://arxiv.org/abs/2305.07644}{arXiv:2305.07644}.

\bibitem{carlini_secret_2019}
N.~Carlini, C.~Liu, U.~Erlingsson, J.~Kos, and D.~Song, ``The {Secret} {Sharer}: {Evaluating} and {Testing} {Unintended} {Memorization} in {Neural} {Networks},'' in {\em 28th {USENIX} {Security} {Symposium}}, (Santa Clara, CA), USENIX Association, Aug. 2019.

\bibitem{hartmann_sok_2023}
V.~Hartmann, A.~Suri, V.~Bindschaedler, D.~Evans, S.~Tople, and R.~West, ``{SoK}: {Memorization} in {General}-{Purpose} {Large} {Language} {Models},'' Oct. 2023.
\newblock \href{http://arxiv.org/abs/2310.18362}{arXiv:2310.18362}.

\bibitem{shokri_membership_2017}
R.~Shokri, M.~Stronati, C.~Song, and V.~Shmatikov, ``Membership {Inference} {Attacks} {Against} {Machine} {Learning} {Models},'' in {\em {IEEE} {Symposium} on {Security} and {Privacy} ({SP})}, (San Jose, CA, USA), IEEE, 2017.

\bibitem{hu_membership_2023}
H.~Hu and J.~Pang, ``Membership {Inference} of {Diffusion} {Models},'' Jan. 2023.
\newblock \href{http://arxiv.org/abs/2301.09956}{arXiv:2301.09956}.

\bibitem{fredrikson_model_2015}
M.~Fredrikson, S.~Jha, and T.~Ristenpart, ``Model {Inversion} {Attacks} that {Exploit} {Confidence} {Information} and {Basic} {Countermeasures},'' in {\em {ACM} {SIGSAC} {Conference} on {Computer} and {Communications} {Security}}, {CCS}, (Denver Colorado USA), ACM, Oct. 2015.

\bibitem{yin_dreaming_2020}
H.~Yin, P.~Molchanov, J.~M. Alvarez, Z.~Li, A.~Mallya, D.~Hoiem, N.~K. Jha, and J.~Kautz, ``Dreaming to {Distill}: {Data}-{Free} {Knowledge} {Transfer} via {DeepInversion},'' in {\em {IEEE}/{CVF} {Conference} on {Computer} {Vision} and {Pattern} {Recognition} ({CVPR})}, (Seattle, WA, USA), IEEE, June 2020.

\bibitem{ghiasi_plug-inversion_2022}
A.~Ghiasi, H.~Kazemi, S.~Reich, C.~Zhu, M.~Goldblum, and T.~Goldstein, ``Plug-{In} {Inversion}: {Model}-{Agnostic} {Inversion} for {Vision} with {Data} {Augmentations},'' Jan. 2022.
\newblock \href{http://arxiv.org/abs/2201.12961}{arXiv:2201.12961}.

\bibitem{collins_editing_2020}
E.~Collins, R.~Bala, B.~Price, and S.~Susstrunk, ``Editing in {Style}: {Uncovering} the {Local} {Semantics} of {GANs},'' in {\em {IEEE}/{CVF} {Conference} on {Computer} {Vision} and {Pattern} {Recognition} ({CVPR})}, (Seattle, WA, USA), IEEE, June 2020.

\bibitem{harkonen_ganspace_2020}
E.~Härkönen, A.~Hertzmann, J.~Lehtinen, and S.~Paris, ``{GANSpace}: {Discovering} {Interpretable} {GAN} {Controls},'' Dec. 2020.
\newblock \href{http://arxiv.org/abs/2004.02546}{arXiv:2004.02546}.

\bibitem{wu_uncovering_2023}
Q.~Wu, Y.~Liu, H.~Zhao, A.~Kale, T.~Bui, T.~Yu, Z.~Lin, Y.~Zhang, and S.~Chang, ``Uncovering the {Disentanglement} {Capability} in {Text}-to-{Image} {Diffusion} {Models},'' in {\em {IEEE}/{CVF} {Conference} on {Computer} {Vision} and {Pattern} {Recognition} ({CVPR})}, (Vancouver, BC, Canada), IEEE, June 2023.

\bibitem{higgins_towards_2018}
I.~Higgins, D.~Amos, D.~Pfau, S.~Racaniere, L.~Matthey, D.~Rezende, and A.~Lerchner, ``Towards a {Definition} of {Disentangled} {Representations},'' Dec. 2018.
\newblock \href{http://arxiv.org/abs/1812.02230}{arXiv:1812.02230}.

\bibitem{yang_disdiff_2023}
T.~Yang, Y.~Wang, Y.~Lv, and N.~Zheng, ``{DisDiff}: {Unsupervised} {Disentanglement} of {Diffusion} {Probabilistic} {Models},'' Oct. 2023.
\newblock \href{http://arxiv.org/abs/2301.13721}{arXiv:2301.13721}.

\bibitem{wang_stylediffusion_2023}
Z.~Wang, L.~Zhao, and W.~Xing, ``{StyleDiffusion}: {Controllable} {Disentangled} {Style} {Transfer} via {Diffusion} {Models},'' in {\em {IEEE}/{CVF} {International} {Conference} on {Computer} {Vision} ({ICCV})}, (Paris, France), IEEE, Oct. 2023.

\bibitem{kwon_diffusion-based_2023}
G.~Kwon and J.~C. Ye, ``Diffusion-based {Image} {Translation} using {Disentangled} {Style} and {Content} {Representation},'' Feb. 2023.
\newblock \href{http://arxiv.org/abs/2209.15264}{arXiv:2209.15264}.

\bibitem{kazemi_style_2018}
H.~Kazemi, S.~M. Iranmanesh, and N.~M. Nasrabadi, ``Style and {Content} {Disentanglement} in {Generative} {Adversarial} {Networks},'' Nov. 2018.
\newblock \href{http://arxiv.org/abs/1811.05621}{arXiv:1811.05621}.

\bibitem{huang_arbitrary_2017}
X.~Huang and S.~Belongie, ``Arbitrary {Style} {Transfer} in {Real}-time with {Adaptive} {Instance} {Normalization},'' July 2017.
\newblock \href{http://arxiv.org/abs/1703.06868}{arXiv:1703.06868}.

\bibitem{kwon_diagonal_2021}
G.~Kwon and J.~C. Ye, ``Diagonal {Attention} and {Style}-based {GAN} for {Content}-{Style} {Disentanglement} in {Image} {Generation} and {Translation},'' July 2021.
\newblock \href{http://arxiv.org/abs/2103.16146}{arXiv:2103.16146}.

\bibitem{xu_drb-gan_2021}
W.~Xu, C.~Long, R.~Wang, and G.~Wang, ``{DRB}-{GAN}: {A} {Dynamic} {ResBlock} {Generative} {Adversarial} {Network} for {Artistic} {Style} {Transfer},'' in {\em {IEEE}/{CVF} {International} {Conference} on {Computer} {Vision} ({ICCV})}, (Montreal, QC, Canada), IEEE, Oct. 2021.

\bibitem{liu_artsy-gan_2018}
H.~Liu, P.~N. Michelini, and D.~Zhu, ``Artsy-{GAN}: {A} style transfer system with improved quality, diversity and performance,'' in {\em 24th {International} {Conference} on {Pattern} {Recognition} ({ICPR})}, (Beijing, China), IEEE, Aug. 2018.

\bibitem{wu_not_2023}
Y.~Wu, Y.~Nakashima, and N.~Garcia, ``Not {Only} {Generative} {Art}: {Stable} {Diffusion} for {Content}-{Style} {Disentanglement} in {Art} {Analysis},'' in {\em {ACM} {International} {Conference} on {Multimedia} {Retrieval}}, (Thessaloniki Greece), ACM, June 2023.

\bibitem{kotovenko_content_2019}
D.~Kotovenko, A.~Sanakoyeu, S.~Lang, and B.~Ommer, ``Content and {Style} {Disentanglement} for {Artistic} {Style} {Transfer},'' in {\em {IEEE}/{CVF} {International} {Conference} on {Computer} {Vision} ({ICCV})}, (Seoul, Korea (South)), IEEE, Oct. 2019.

\bibitem{shan_glaze_2023}
S.~Shan, J.~Cryan, E.~Wenger, H.~Zheng, R.~Hanocka, and B.~Y. Zhao, ``Glaze: {Protecting} {Artists} from {Style} {Mimicry} by {Text}-to-{Image} {Models},'' in {\em 32nd {USENIX} {Security} {Symposium}}, (Anaheim, CA), USENIX Association, Aug. 2023.

\bibitem{liang_adversarial_2023}
C.~Liang, X.~Wu, Y.~Hua, J.~Zhang, Y.~Xue, T.~Song, Z.~Xue, R.~Ma, and H.~Guan, ``Adversarial {Example} {Does} {Good}: {Preventing} {Painting} {Imitation} from {Diffusion} {Models} via {Adversarial} {Examples},'' June 2023.
\newblock \href{http://arxiv.org/abs/2302.04578}{arXiv:2302.04578}.

\bibitem{liang_mist_2023}
C.~Liang and X.~Wu, ``Mist: {Towards} {Improved} {Adversarial} {Examples} for {Diffusion} {Models},'' May 2023.
\newblock \href{http://arxiv.org/abs/2305.12683}{arXiv:2305.12683}.

\bibitem{ye_duaw_2023}
X.~Ye, H.~Huang, J.~An, and Y.~Wang, ``{DUAW}: {Data}-free {Universal} {Adversarial} {Watermark} against {Stable} {Diffusion} {Customization},'' Aug. 2023.
\newblock \href{http://arxiv.org/abs/2308.09889}{arXiv:2308.09889}.

\bibitem{zhao_unlearnable_2023}
Z.~Zhao, J.~Duan, X.~Hu, K.~Xu, C.~Wang, R.~Zhang, Z.~Du, Q.~Guo, and Y.~Chen, ``Unlearnable {Examples} for {Diffusion} {Models}: {Protect} {Data} from {Unauthorized} {Exploitation},'' June 2023.
\newblock \href{http://arxiv.org/abs/2306.01902}{arXiv:2306.01902}.

\bibitem{chen_editshield_2023}
R.~Chen, H.~Jin, J.~Chen, and L.~Sun, ``{EditShield}: {Protecting} {Unauthorized} {Image} {Editing} by {Instruction}-guided {Diffusion} {Models},'' Nov. 2023.
\newblock \href{http://arxiv.org/abs/2311.12066}{arXiv:2311.12066}.

\bibitem{salman_raising_2023}
H.~Salman, A.~Khaddaj, G.~Leclerc, A.~Ilyas, and A.~Madry, ``Raising the {Cost} of {Malicious} {AI}-{Powered} {Image} {Editing},'' Feb. 2023.
\newblock \href{http://arxiv.org/abs/2302.06588}{arXiv:2302.06588}.

\bibitem{shan_prompt-specific_2023}
S.~Shan, W.~Ding, J.~Passananti, H.~Zheng, and B.~Y. Zhao, ``Prompt-{Specific} {Poisoning} {Attacks} on {Text}-to-{Image} {Generative} {Models},'' Oct. 2023.
\newblock \href{http://arxiv.org/abs/2310.13828}{arXiv:2310.13828}.

\bibitem{zheng_understanding_2023}
B.~Zheng, C.~Liang, X.~Wu, and Y.~Liu, ``Understanding and {Improving} {Adversarial} {Attacks} on {Latent} {Diffusion} {Model},'' Oct. 2023.
\newblock \href{http://arxiv.org/abs/2310.04687}{arXiv:2310.04687}.

\bibitem{van_le_anti-dreambooth_2023}
T.~Van~Le, H.~Phung, T.~H. Nguyen, Q.~Dao, N.~Tran, and A.~Tran, ``Anti-{DreamBooth}: {Protecting} users from personalized text-to-image synthesis,'' Oct. 2023.
\newblock \href{http://arxiv.org/abs/2303.15433}{arXiv:2303.15433}.

\bibitem{liu_metacloak_2024}
Y.~Liu, C.~Fan, Y.~Dai, X.~Chen, P.~Zhou, and L.~Sun, ``{MetaCloak}: {Preventing} {Unauthorized} {Subject}-driven {Text}-to-image {Diffusion}-based {Synthesis} via {Meta}-learning,'' in {\em {IEEE}/{CVF} {Conference} on {Computer} {Vision} and {Pattern} {Recognition} ({CVPR})}, (Seattle WA, USA), IEEE, July 2024.

\bibitem{li_mitigate_2024}
C.~Li, D.~Chen, Y.~Zhang, and P.~A. Beerel, ``Mitigate {Replication} and {Copying} in {Diffusion} {Models} with {Generalized} {Caption} and {Dual} {Fusion} {Enhancement},'' Jan. 2024.
\newblock \href{http://arxiv.org/abs/2309.07254}{arXiv:2309.07254}.

\bibitem{kong_data_2023}
Z.~Kong and K.~Chaudhuri, ``Data {Redaction} from {Pre}-trained {GANs},'' in {\em {IEEE} {Conference} on {Secure} and {Trustworthy} {Machine} {Learning} ({SaTML})}, (Raleigh, NC, USA), IEEE, Feb. 2023.

\bibitem{schramowski_safe_2023}
P.~Schramowski, M.~Brack, B.~Deiseroth, and K.~Kersting, ``Safe {Latent} {Diffusion}: {Mitigating} {Inappropriate} {Degeneration} in {Diffusion} {Models},'' Apr. 2023.
\newblock \href{http://arxiv.org/abs/2211.05105}{arXiv:2211.05105}.

\bibitem{gandikota_unified_2024}
R.~Gandikota, H.~Orgad, Y.~Belinkov, J.~Materzyńska, and D.~Bau, ``Unified {Concept} {Editing} in {Diffusion} {Models},'' in {\em {IEEE}/{CVF} {Winter} {Conference} on {Applications} of {Computer} {Vision} ({WACV})}, (Waikoloa, HI, USA), IEEE, Jan. 2024.

\bibitem{kumari_ablating_2023}
N.~Kumari, B.~Zhang, S.-Y. Wang, E.~Shechtman, R.~Zhang, and J.-Y. Zhu, ``Ablating {Concepts} in {Text}-to-{Image} {Diffusion} {Models},'' May 2023.
\newblock \href{http://arxiv.org/abs/2303.13516}{arXiv:2303.13516}.

\bibitem{zhang_forget-me-not_2023}
E.~Zhang, K.~Wang, X.~Xu, Z.~Wang, and H.~Shi, ``Forget-{Me}-{Not}: {Learning} to {Forget} in {Text}-to-{Image} {Diffusion} {Models},'' Mar. 2023.
\newblock \href{http://arxiv.org/abs/2303.17591}{arXiv:2303.17591}.

\bibitem{eldan_whos_2023}
R.~Eldan and M.~Russinovich, ``Who's {Harry} {Potter}? {Approximate} {Unlearning} in {LLMs},'' Oct. 2023.
\newblock \href{http://arxiv.org/abs/2310.02238}{arXiv:2310.02238}.

\bibitem{liu_rethinking_2024}
S.~Liu, Y.~Yao, J.~Jia, S.~Casper, N.~Baracaldo, P.~Hase, X.~Xu, Y.~Yao, H.~Li, K.~R. Varshney, M.~Bansal, S.~Koyejo, and Y.~Liu, ``Rethinking {Machine} {Unlearning} for {Large} {Language} {Models},'' Feb. 2024.
\newblock \href{http://arxiv.org/abs/2402.08787}{arXiv:2402.08787}.

\bibitem{noauthor_openai_2023}
``{OpenAI} {Moderation},'' 2023.
\newblock https://platform.openai.com/docs/guides/moderation (accessed 2024-01-31).

\bibitem{hanu_detoxify_2020}
L.~Hanu and {Unitary Team}, ``Detoxify,'' Nov. 2020.
\newblock https://github.com/unitaryai/detoxify (accessed 204-01-31).

\bibitem{cui_diffusionshield_2023}
Y.~Cui, J.~Ren, H.~Xu, P.~He, H.~Liu, L.~Sun, Y.~Xing, and J.~Tang, ``{DiffusionShield}: {A} {Watermark} for {Copyright} {Protection} against {Generative} {Diffusion} {Models},'' Oct. 2023.
\newblock \href{http://arxiv.org/abs/2306.04642}{arXiv:2306.04642}.

\bibitem{zhang_editguard_2023}
X.~Zhang, R.~Li, J.~Yu, Y.~Xu, W.~Li, and J.~Zhang, ``{EditGuard}: {Versatile} {Image} {Watermarking} for {Tamper} {Localization} and {Copyright} {Protection},'' Dec. 2023.
\newblock \href{http://arxiv.org/abs/2312.08883}{arXiv:2312.08883}.

\bibitem{hayes_generating_2017}
J.~Hayes and G.~Danezis, ``Generating steganographic images via adversarial training,'' in {\em Advances in {Neural} {Information} {Processing} {Systems}}, vol.~30, Curran Associates, Inc., 2017.

\bibitem{ma_generative_2023}
Y.~Ma, Z.~Zhao, X.~He, Z.~Li, M.~Backes, and Y.~Zhang, ``Generative {Watermarking} {Against} {Unauthorized} {Subject}-{Driven} {Image} {Synthesis},'' June 2023.
\newblock \href{http://arxiv.org/abs/2306.07754}{arXiv:2306.07754}.

\bibitem{cui_ft-shield_2023}
Y.~Cui, J.~Ren, Y.~Lin, H.~Xu, P.~He, Y.~Xing, W.~Fan, H.~Liu, and J.~Tang, ``{FT}-{Shield}: {A} {Watermark} {Against} {Unauthorized} {Fine}-tuning in {Text}-to-{Image} {Diffusion} {Models},'' Oct. 2023.
\newblock \href{http://arxiv.org/abs/2310.02401}{arXiv:2310.02401}.

\bibitem{tan_somewhat_2023}
M.~Tan, T.~Wang, and S.~Jha, ``A {Somewhat} {Robust} {Image} {Watermark} against {Diffusion}-based {Editing} {Models},'' Nov. 2023.
\newblock \href{http://arxiv.org/abs/2311.13713}{arXiv:2311.13713}.

\bibitem{liu_detecting_2024}
C.~Liu, J.~Zhang, T.~Zhang, X.~Yang, W.~Zhang, and N.~Yu, ``Detecting {Voice} {Cloning} {Attacks} via {Timbre} {Watermarking},'' in {\em Network and {Distributed} {System} {Security} {Symposium}}, (San Diego, CA, USA), Internet Society, 2024.

\bibitem{feng_catch_2023}
W.~Feng, J.~He, J.~Zhang, T.~Zhang, W.~Zhou, W.~Zhang, and N.~Yu, ``Catch {You} {Everything} {Everywhere}: {Guarding} {Textual} {Inversion} via {Concept} {Watermarking},'' Sept. 2023.
\newblock \href{http://arxiv.org/abs/2309.05940}{arXiv:2309.05940}.

\bibitem{wang_evaluating_2023}
S.-Y. Wang, A.~A. Efros, J.-Y. Zhu, and R.~Zhang, ``Evaluating {Data} {Attribution} for {Text}-to-{Image} {Models},'' Aug. 2023.
\newblock \href{http://arxiv.org/abs/2306.09345}{arXiv:2306.09345}.

\bibitem{dai_training_2023}
Z.~Dai and D.~K. Gifford, ``Training {Data} {Attribution} for {Diffusion} {Models},'' June 2023.
\newblock \href{http://arxiv.org/abs/2306.02174}{arXiv:2306.02174}.

\bibitem{tsai_ring--bell_2023}
Y.-L. Tsai, C.-Y. Hsu, C.~Xie, C.-H. Lin, J.-Y. Chen, B.~Li, P.-Y. Chen, C.-M. Yu, and C.-Y. Huang, ``Ring-{A}-{Bell}! {How} {Reliable} are {Concept} {Removal} {Methods} for {Diffusion} {Models}?,'' Oct. 2023.
\newblock \href{http://arxiv.org/abs/2310.10012}{arXiv:2310.10012}.

\bibitem{vyas_provable_2023}
N.~Vyas, S.~Kakade, and B.~Barak, ``Provable {Copyright} {Protection} for {Generative} {Models},'' Feb. 2023.
\newblock \href{http://arxiv.org/abs/2302.10870}{arXiv:2302.10870}.

\bibitem{lu_specialist_2023}
H.~Lu, H.~Tunanyan, K.~Wang, S.~Navasardyan, Z.~Wang, and H.~Shi, ``Specialist {Diffusion}: {Plug}-and-{Play} {Sample}-{Efficient} {Fine}-{Tuning} of {Text}-to-{Image} {Diffusion} {Models} to {Learn} {Any} {Unseen} {Style},'' in {\em {IEEE}/{CVF} {Conference} on {Computer} {Vision} and {Pattern} {Recognition} ({CVPR})}, (Vancouver, BC, Canada), IEEE, June 2023.

\bibitem{kandpal_large_2023}
N.~Kandpal, H.~Deng, A.~Roberts, E.~Wallace, and C.~Raffel, ``Large {Language} {Models} {Struggle} to {Learn} {Long}-{Tail} {Knowledge},'' in {\em 40th {International} {Conference} on {Machine} {Learning} ({ICML})}, PMLR, July 2023.

\bibitem{webster_-duplication_2023}
R.~Webster, J.~Rabin, L.~Simon, and F.~Jurie, ``On the {De}-duplication of {LAION}-{2B},'' Mar. 2023.
\newblock \href{http://arxiv.org/abs/2303.12733}{arXiv:2303.12733}.

\bibitem{li_blip_2022}
J.~Li, D.~Li, C.~Xiong, and S.~Hoi, ``{BLIP}: {Bootstrapping} {Language}-{Image} {Pre}-training for {Unified} {Vision}-{Language} {Understanding} and {Generation},'' Feb. 2022.
\newblock \href{http://arxiv.org/abs/2201.12086}{arXiv:2201.12086}.

\bibitem{goodfellow_explaining_2015}
I.~J. Goodfellow, J.~Shlens, and C.~Szegedy, ``Explaining and {Harnessing} {Adversarial} {Examples},'' Mar. 2015.
\newblock \href{http://arxiv.org/abs/1412.6572}{arXiv:1412.6572}.

\bibitem{rando_red-teaming_2022}
J.~Rando, D.~Paleka, D.~Lindner, L.~Heim, and F.~Tramèr, ``Red-{Teaming} the {Stable} {Diffusion} {Safety} {Filter},'' Nov. 2022.
\newblock \href{http://arxiv.org/abs/2210.04610}{arXiv:2210.04610}.

\bibitem{jang_knowledge_2022}
J.~Jang, D.~Yoon, S.~Yang, S.~Cha, M.~Lee, L.~Logeswaran, and M.~Seo, ``Knowledge {Unlearning} for {Mitigating} {Privacy} {Risks} in {Language} {Models},'' Dec. 2022.
\newblock \href{http://arxiv.org/abs/2210.01504}{arXiv:2210.01504}.

\bibitem{wu_depn_2023}
X.~Wu, J.~Li, M.~Xu, W.~Dong, S.~Wu, C.~Bian, and D.~Xiong, ``{DEPN}: {Detecting} and {Editing} {Privacy} {Neurons} in {Pretrained} {Language} {Models},'' Dec. 2023.
\newblock \href{http://arxiv.org/abs/2310.20138}{arXiv:2310.20138}.

\bibitem{lu_quark_2022}
X.~Lu, S.~Welleck, J.~Hessel, L.~Jiang, L.~Qin, P.~West, P.~Ammanabrolu, and Y.~Choi, ``{QUARK}: {Controllable} {Text} {Generation} with {Reinforced} {Unlearning},'' in {\em Advances in {Neural} {Information} {Processing} {Systems}}, vol.~35, Curran Associates, Inc., 2022.

\bibitem{navas_dwt-dct-svd_2008}
K.~A. Navas, M.~C. Ajay, M.~Lekshmi, T.~S. Archana, and M.~Sasikumar, ``{DWT}-{DCT}-{SVD} based watermarking,'' in {\em 3rd {International} {Conference} on {Communication} {Systems} {Software} and {Middleware} and {Workshops} ({COMSWARE})}, (Bangalore, India), IEEE, Jan. 2008.

\bibitem{zhu_hidden_2018}
J.~Zhu, R.~Kaplan, J.~Johnson, and L.~Fei-Fei, ``{HiDDeN}: {Hiding} {Data} {With} {Deep} {Networks},'' in {\em European {Conference} on {Computer} {Vision} ({ECCV})}, (Munich, Germany), Springer International Publishing, 2018.

\bibitem{yu_artificial_2021}
N.~Yu, V.~Skripniuk, S.~Abdelnabi, and M.~Fritz, ``Artificial {Fingerprinting} for {Generative} {Models}: {Rooting} {Deepfake} {Attribution} in {Training} {Data},'' in {\em {IEEE}/{CVF} {International} {Conference} on {Computer} {Vision} ({ICCV})}, (Montreal, QC, Canada), IEEE, Oct. 2021.

\bibitem{caron_emerging_2021}
M.~Caron, H.~Touvron, I.~Misra, H.~Jegou, J.~Mairal, P.~Bojanowski, and A.~Joulin, ``Emerging {Properties} in {Self}-{Supervised} {Vision} {Transformers},'' in {\em {IEEE}/{CVF} {International} {Conference} on {Computer} {Vision} ({ICCV})}, (Montreal, QC, Canada), IEEE, Oct. 2021.

\bibitem{dosovitskiy_image_2021}
A.~Dosovitskiy, L.~Beyer, A.~Kolesnikov, D.~Weissenborn, X.~Zhai, T.~Unterthiner, M.~Dehghani, M.~Minderer, G.~Heigold, S.~Gelly, J.~Uszkoreit, and N.~Houlsby, ``An {Image} is {Worth} 16x16 {Words}: {Transformers} for {Image} {Recognition} at {Scale},'' June 2021.
\newblock \href{http://arxiv.org/abs/2010.11929}{arXiv:2010.11929}.

\bibitem{ruta_aladin_2021}
D.~Ruta, S.~Motiian, B.~Faieta, Z.~Lin, H.~Jin, A.~Filipkowski, A.~Gilbert, and J.~Collomosse, ``{ALADIN}: {All} {Layer} {Adaptive} {Instance} {Normalization} for {Fine}-grained {Style} {Similarity},'' in {\em {IEEE}/{CVF} {International} {Conference} on {Computer} {Vision} ({ICCV})}, (Montreal, QC, Canada), IEEE, Oct. 2021.

\bibitem{oord_representation_2019}
A.~v.~d. Oord, Y.~Li, and O.~Vinyals, ``Representation {Learning} with {Contrastive} {Predictive} {Coding},'' Jan. 2019.
\newblock \href{http://arxiv.org/abs/1807.03748}{arXiv:1807.03748}.

\bibitem{wu_how_2023}
G.~Wu, ``How to {Know} if {Your} {Images} {Trained} an {AI} {Model} (and {How} to {Opt} {Out}),'' Jan. 2023.
\newblock https://www.makeuseof.com/how-to-know-images-trained-ai-art-generator/ (accessed 204-01-31).

\bibitem{wiggers_deviantart_2022}
K.~Wiggers, ``{DeviantArt} provides a way for artists to opt out of {AI} art generators,'' Nov. 2022.
\newblock https://techcrunch.com/2022/11/11/deviantart-provides-a-way-for-artists-to-opt-out-of-ai-art-generators/ (accessed 2024-01-31).

\bibitem{bogle_new_2023}
A.~Bogle, ``New {York} {Times}, {CNN} and {Australia}’s {ABC} block {OpenAI}’s {GPTBot} web crawler from accessing content,'' Aug. 2023.
\newblock https://www.theguardian.com/technology/2023/aug/25/new-york-times-cnn-and-abc-block-openais-gptbot-web-crawler-from-scraping-content (accessed 2024-01-31).

\bibitem{marcus_generative_2024}
G.~Marcus and R.~Southen, ``Generative {AI} {Has} a {Visual} {Plagiarism} {Problem},'' June 2024.
\newblock https://spectrum.ieee.org/midjourney-copyright (accessed 2024-01-31).

\bibitem{knott_generative_2023}
A.~Knott, D.~Pedreschi, R.~Chatila, T.~Chakraborti, S.~Leavy, R.~Baeza-Yates, D.~Eyers, A.~Trotman, P.~D. Teal, P.~Biecek, S.~Russell, and Y.~Bengio, ``Generative {AI} models should include detection mechanisms as a condition for public release,'' {\em Ethics and Information Technology}, vol.~25, Oct. 2023.

\bibitem{chen_gmail_2019}
M.~X. Chen, B.~N. Lee, G.~Bansal, Y.~Cao, S.~Zhang, J.~Lu, J.~Tsay, Y.~Wang, A.~M. Dai, Z.~Chen, T.~Sohn, and Y.~Wu, ``Gmail {Smart} {Compose}: {Real}-{Time} {Assisted} {Writing},'' in {\em 25th {ACM} {SIGKDD} {International} {Conference} on {Knowledge} {Discovery} \& {Data} {Mining}}, (Anchorage AK USA), ACM, July 2019.

\bibitem{liu_geom-erasing_2023}
Z.~Liu, K.~Chen, Y.~Zhang, J.~Han, L.~Hong, H.~Xu, Z.~Li, D.-Y. Yeung, and J.~Kwok, ``Geom-{Erasing}: {Geometry}-{Driven} {Removal} of {Implicit} {Concept} in {Diffusion} {Models},'' Oct. 2023.
\newblock \href{http://arxiv.org/abs/2310.05873}{arXiv:2310.05873}.

\bibitem{ho_classifier-free_2022}
J.~Ho and T.~Salimans, ``Classifier-{Free} {Diffusion} {Guidance},'' July 2022.
\newblock \href{http://arxiv.org/abs/2207.12598}{arXiv:2207.12598}.

\bibitem{wu_erasediff_2024}
J.~Wu, T.~Le, M.~Hayat, and M.~Harandi, ``{EraseDiff}: {Erasing} {Data} {Influence} in {Diffusion} {Models},'' Feb. 2024.
\newblock \href{http://arxiv.org/abs/2401.05779}{arXiv:2401.05779}.

\bibitem{wu_towards_2023}
R.~Wu, Y.~Wang, H.~Shi, Z.~Yu, Y.~Wu, and D.~Liang, ``Towards {Prompt}-robust {Face} {Privacy} {Protection} via {Adversarial} {Decoupling} {Augmentation} {Framework},'' May 2023.
\newblock \href{http://arxiv.org/abs/2305.03980}{arXiv:2305.03980}.

\bibitem{ruiz_disrupting_2020}
N.~Ruiz, S.~A. Bargal, and S.~Sclaroff, ``Disrupting {Deepfakes}: {Adversarial} {Attacks} {Against} {Conditional} {Image} {Translation} {Networks} and {Facial} {Manipulation} {Systems},'' in {\em European {Conference} on {Computer} {Vision} ({ECCV}) {Workshops}}, (Glasgow, UK), Springer International Publishing, 2020.

\bibitem{dong_restricted_2023}
J.~Dong, Y.~Wang, J.~Lai, and X.~Xie, ``Restricted {Black}-{Box} {Adversarial} {Attack} {Against} {DeepFake} {Face} {Swapping},'' {\em IEEE Transactions on Information Forensics and Security}, vol.~18, 2023.

\bibitem{ruiz_practical_2023}
N.~Ruiz, S.~A. Bargal, C.~Xie, and S.~Sclaroff, ``Practical {Disruption} of {Image} {Translation} {Deepfake} {Networks},'' in {\em {AAAI} {Conference} on {Artificial} {Intelligence}}, vol.~37, (Washington DC, USA), June 2023.

\bibitem{wang_faketagger_2021}
R.~Wang, F.~Juefei-Xu, M.~Luo, Y.~Liu, and L.~Wang, ``{FakeTagger}: {Robust} {Safeguards} against {DeepFake} {Dissemination} via {Provenance} {Tracking},'' in {\em 29th {ACM} {International} {Conference} on {Multimedia}}, (Virtual Event China), ACM, Oct. 2021.

\bibitem{wu_sepmark_2023}
X.~Wu, X.~Liao, and B.~Ou, ``{SepMark}: {Deep} {Separable} {Watermarking} for {Unified} {Source} {Tracing} and {Deepfake} {Detection},'' in {\em 31st {ACM} {International} {Conference} on {Multimedia}}, (Ottawa ON Canada), ACM, Oct. 2023.

\bibitem{tang_science_2024}
R.~Tang, Y.-N. Chuang, and X.~Hu, ``The {Science} of {Detecting} {LLM}-{Generated} {Text},'' {\em Communications of the ACM}, vol.~67, Apr. 2024.

\bibitem{liu_detecting_2022}
B.~Liu, F.~Yang, X.~Bi, B.~Xiao, W.~Li, and X.~Gao, ``Detecting {Generated} {Images} by {Real} {Images},'' in {\em European {Conference} on {Computer} {Vision} ({ECCV})}, (Tel Aviv, Israel), Springer Nature Switzerland, 2022.

\bibitem{sadasivan_can_2024}
V.~S. Sadasivan, A.~Kumar, S.~Balasubramanian, W.~Wang, and S.~Feizi, ``Can {AI}-{Generated} {Text} be {Reliably} {Detected}?,'' Feb. 2024.
\newblock \href{http://arxiv.org/abs/2303.11156}{arXiv:2303.11156}.

\bibitem{weber-wulff_testing_2023}
D.~Weber-Wulff, A.~Anohina-Naumeca, S.~Bjelobaba, T.~Foltýnek, J.~Guerrero-Dib, O.~Popoola, P.~Šigut, and L.~Waddington, ``Testing of detection tools for {AI}-generated text,'' {\em International Journal for Educational Integrity}, vol.~19, Dec. 2023.

\bibitem{corvi_detection_2023}
R.~Corvi, D.~Cozzolino, G.~Zingarini, G.~Poggi, K.~Nagano, and L.~Verdoliva, ``On {The} {Detection} of {Synthetic} {Images} {Generated} by {Diffusion} {Models},'' in {\em {IEEE} {International} {Conference} on {Acoustics}, {Speech} and {Signal} {Processing} ({ICASSP})}, (Rhodes Island, Greece), IEEE, June 2023.

\bibitem{zhang_udh_2020}
C.~Zhang, P.~Benz, A.~Karjauv, G.~Sun, and I.~S. Kweon, ``{UDH}: {Universal} {Deep} {Hiding} for {Steganography}, {Watermarking}, and {Light} {Field} {Messaging},'' in {\em Advances in {Neural} {Information} {Processing} {Systems}}, vol.~33, Curran Associates, Inc., 2020.

\bibitem{jiang_evading_2023}
Z.~Jiang, J.~Zhang, and N.~Z. Gong, ``Evading {Watermark} based {Detection} of {AI}-{Generated} {Content},'' Nov. 2023.
\newblock \href{http://arxiv.org/abs/2305.03807}{arXiv:2305.03807}.

\bibitem{lnu_artists_2023}
D.~Lnu, ``Artists enable {AI} art - shouldn't they be compensated?,'' Feb. 2023.

\bibitem{lemley_fair_2021}
M.~A. Lemley and B.~Casey, ``Fair {Learning},'' {\em Texas Law Review}, vol.~99, no.~4, 2021.

\bibitem{roose_ai-generated_2022}
K.~Roose, ``{AI}-{Generated} {Art} {Won} a {Prize}. {Artists} {Aren}’t {Happy}.,'' Feb. 2022.
\newblock https://www.nytimes.com/2022/09/02/technology/ai-artificial-intelligence-artists.html.

\bibitem{brodkin_us_2023}
J.~Brodkin, ``{US} judge: {Art} created solely by artificial intelligence cannot be copyrighted,'' Aug. 2023.
\newblock https://arstechnica.com/tech-policy/2023/08/us-judge-art-created-solely-by-artificial-intelligence-cannot-be-copyrighted/.

\bibitem{meng_magnet_2017}
D.~Meng and H.~Chen, ``{MagNet}: {A} {Two}-{Pronged} {Defense} against {Adversarial} {Examples},'' in {\em {ACM} {SIGSAC} {Conference} on {Computer} and {Communications} {Security} ({CCS})}, (Dallas Texas USA), ACM, Oct. 2017.

\bibitem{qin_destruction-restoration_2023}
T.~Qin, X.~Gao, J.~Zhao, and K.~Ye, ``Destruction-{Restoration} {Suppresses} {Data} {Protection} {Perturbations} against {Diffusion} {Models},'' in {\em 35th {International} {Conference} on {Tools} with {Artificial} {Intelligence} ({ICTAI})}, (Atlanta, GA, USA), IEEE, Nov. 2023.

\end{thebibliography}
